\documentclass[preprint]{imsart}

\RequirePackage[OT1]{fontenc}
\RequirePackage[numbers]{natbib}

\usepackage{amsthm,amsmath}
\usepackage{amsbsy}

\input xy
\xyoption{all}

\usepackage{algorithm}
\usepackage{algpseudocode}

\usepackage{epsf}
\usepackage{epsfig}

\usepackage{enumitem}

\usepackage{geometry}
\newcommand{\bfX}{\boldsymbol{X}}
\newcommand{\bfy}{\boldsymbol{y}}
\newcommand{\bfc}{\boldsymbol{c}}

\newtheorem{theorem}{Theorem}

\begin{document}

\begin{frontmatter}

\title{Using permutations to detect, quantify and correct for confounding in machine learning predictions}
\runtitle{Permutations to detect, quantify and correct for confounding in machine learning}



\author{\fnms{} \snm{Elias Chaibub Neto$^{1}$}\ead[label=e1]{$^\ast$ elias.chaibub.neto@sagebase.org, $^1$ Sage Bionetworks}}
\address{\printead{e1}}



\runauthor{Chaibub Neto E.}

\begin{abstract}
Clinical machine learning applications are often plagued with confounders that are clinically irrelevant, but can still artificially boost the predictive performance of the algorithms. Confounding is especially problematic in mobile health studies run ``in the wild", where it is challenging to balance the demographic characteristics of participants that self select to enter the study. An effective approach to remove the influence of confounders is to match samples in order to improve the balance in the data. The caveat is that we end-up with a smaller number of participants to train and evaluate the machine learning algorithm. Alternative confounding adjustment methods that make more efficient use of the data (e.g., inverse probability weighting) usually rely on modeling assumptions, and it is unclear how robust these methods are to violations of these assumptions. Here, rather than proposing a new approach to prevent/reduce the learning of confounding signals by a machine learning algorithm, we develop novel statistical tools to detect, quantify and correct for the influence of observed confounders. Our tools are based on restricted and standard permutation approaches and can be used to evaluate how well a confounding adjustment method is actually working. We use restricted permutations to test if an algorithm has learned disease signal in the presence of confounding signal, and to develop a novel statistical test to detect confounding learning per se. Furthermore, we prove that restricted permutations provide an alternative method to compute partial correlations, and use this result as a motivation to develop a novel approach to estimate the corrected predictive performance of a learner. We evaluate the statistical properties of our methods in simulation studies.
\end{abstract}





\end{frontmatter}

\section{Introduction}

Machine learning algorithms have been increasingly used as diagnostic and prognostic tools in biomedical research. In the emerging field of mobile health, machine learning is especially well positioned to impact clinical research, as the widespread availability of smartphones and other health tracking devices generates high volumes of sensor data that can be readily harnessed to train learners. In clinical applications, gender, age, and other demographic characteristics of the study participants often play the role of confounders. Confounding is particularly prevalent in mobile health studies run ``in the wild" (i.e., under uncontrolled conditions outside clinical and laboratory settings) where we have little control over the demographic and clinical characteristics of the cohort of participants that self-select to enter the study.

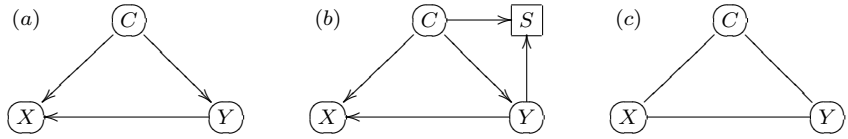
\begin{figure}[!h]
$$
\xymatrix@-0.0pc{
(a) & *+[F-:<10pt>]{C} \ar[dl] \ar[dr] & & (b) & *+[F-:<10pt>]{C} \ar[dl] \ar[dr] \ar[r] & *+[F]{S}  & (c) & *+[F-:<10pt>]{C} \ar@{-}[dl] \ar@{-}[dr] \\
*+[F-:<10pt>]{X} & & *+[F-:<10pt>]{Y} \ar[ll] & *+[F-:<10pt>]{X} & & *+[F-:<10pt>]{Y} \ar[ll] \ar[u] & *+[F-:<10pt>]{X} & & *+[F-:<10pt>]{Y} \ar@{-}[ll] \\
}
$$
\caption{Consider a mobile health diagnostic system for Parkinson's disease (PD), built using features extracted from accelerometer sensor data. Let $Y$, $C$, and $X$ represent, respectively, the disease status, a confounder (such as age or gender), and a accelerometer feature. The causal diagram in panel a describes our assumptions about the causal relations between these variables, in the case where $C$ represents the gender confounder. Because PD patients experience difficulty to walk, their acceleration patterns tend to be distinct from control participants. Hence, we assume that disease status has a causal effect on the feature ($Y \rightarrow X$). We also assume a causal influence of gender on the feature ($C \rightarrow X$), since males on average are taller than females and taller participants tend to have larger step sizes, and different heel strikes and acceleration patterns than shorter participants. Finally, because gender is a risk factor for PD (the disease is more prevalent in males than females) we also assume a causal influence of gender on disease status ($C \rightarrow Y$). The causal diagram in panel b represents our assumptions for the age confounder. As before, we assume that $Y \rightarrow X$,  $C \rightarrow X$ (since older subjects tend to move slower than younger ones), and that $C \rightarrow Y$ (since older age is also a risk factor for PD). Now, let $S$ represent a binary selection variable indicating whether a potential participant has enrolled or not in the study (i.e., $S = 1$ if the person enrolled, and $S = 0$ otherwise). We assume that age influences enrollment in the study ($C \rightarrow S$) because younger people tend to be more technology savvy than older people (and tech savvy people are more likely to enroll in mobile health studies). We also assume that disease status influences enrollment ($Y \rightarrow S$), since patients suffering from a disease are usually more motivated to enroll in a study than controls. The squared frame around $S$ in panel b indicates that the analysis is conditional on the participants that actually enrolled in the study (i.e, is conditional on $S = 1$). Observe that in both panels a and b, $C$ is a confounder of the $Y \rightarrow X$ relation, since it is responsible for spurious associations between $X$ and $Y$ via the backdoor path $X \leftarrow C \rightarrow Y$ in panel a, and the backdoor paths $X \leftarrow C \rightarrow Y$ and $X \leftarrow C \rightarrow S \leftarrow Y$ in panel b (note that conditional on $S = 1$ this last backdoor path is open through the collider $C \rightarrow S \leftarrow Y$). Finally, panel c represents the undirected dependency graph (UDG) associated with both causal diagrams in panels a and b (an undirected edge between two nodes means that the variables are associated even when we condition on the remaining variables). The UDG represents the ``association shadows" casted by the causal model underlying the data generation process.}
\label{fig:causaldiagram}
\end{figure}

In the context of predictive modeling, we define a confounder as a variable that causes spurious associations between the features and response variable. We subscribe to Pearl's causal inference framework (Pearl 2000), where potential confounders are identified from a causal diagram describing our qualitative assumptions about the causal relationships between the features, response, and potential confounder variables. Figure \ref{fig:causaldiagram} provides a couple of examples in the context of a Parkinson's disease diagnostic classification problem.

In machine learning applications, the spurious associations generated by confounders can artificially boost the predictive ability of the algorithms. As a concrete example, consider a diagnostic system trained to classify disease cases and healthy controls. Suppose that the disease has a similar prevalence in males and females in the general population, but due to a potentially unknown selection mechanism the data available to develop the classifier is biased in relation to the general population, with most case subjects being males, and most control subjects being female. Furthermore, suppose that the features used to build the classifier are able to efficiently detect gender related signals but not disease related signals. In such a situation, the classifier might achieve excellent classification performance when trained and evaluated on training/test splits of the biased sample since it is able to differentiate males from females (rather than sick and healthy subjects). Such classifier would, nonetheless, perform poorly if deployed in the population of interest (i.e., the general population) where the joint distribution of disease status and gender is shifted relative to the data used to develop the system.

Confounding adjustment is an active area of research in machine learning. Because any variable that confounds the feature/response relationship has to be associated with both the features and the response, most of the methods proposed in the literature can be divided into approaches that either: (i) remove the association between the confounder and the features; or (ii) remove the association between the confounder and the response. A canonical example of the first approach, often denoted the ``adjusted features" approach, is to separately regress each feature on the confounders, and use the residuals of these regressions as the predictors in the machine learning algorithm. A standard example of the second approach is to match subjects from the biased development sample in order to obtain a more balanced subsample of the original data. (The caveat is that we end-up with a smaller number of participants to train and evaluate the machine learning algorithm, and, in highly unbalanced situations, we might end up having to exclude most of the participants from the analyses.) Alternative methods that make more efficient use of the data include inverse probability weighting approaches (Linn et al. 2016, Rao et al. 2017), that weight the training samples in order to make model training better tailored to the population of interest. Approximate inverse probability weighting algorithms have also been proposed in Linn et al. (2016), which essentially use the sample weights to over-sample the biased training set in order to obtain an artificially augmented training set, where the association between the confounders and the response is removed (or at least reduced). Other approaches that do not fall into categories (i) or (ii) have also been proposed. For instance, Li et al. (2011) developed a penalized support vector machine algorithm that favors solutions based on features that do not correlate with the confounder, while Landeiro and Culotta (2016) employ backdoor adjustment in order to obtain classifiers that are robust to shifts in the confounding/response association from training to test data.

In this paper, rather than proposing yet another method to prevent the learning of confounding signals, we develop novel statistical methods to detect, quantify and correct for the influence of observed confounders.

Following Rao et al. (2017), we adopt restricted permutations (Good 2000) to test if an algorithm is actually learning the response signal in addition to learning the signal from the clinically irrelevant confounders\footnote{Note that we refer to a confounder as ``clinically irrelevant" when we are interested in quantifying the amount of response signal learned by an algorithm. In this sense, confounders, such as age, that are certainly relevant in many clinical contexts are still irrelevant for our goal.}. The key idea is to shuffle the response data within the levels of a categorical/ordinal confounder in order to destroy the direct association between the response and the features while still preserving the indirect association mediated by the confounder. Note that while, in theory, we can only perform the restricted permutations using categorical/ordinal confounders, in practice we can discretize and evaluate continuous confounders as well. Of course, if the discretization is too coarse the discretized confounder might not be able to fully capture the association between the confounder and the response, and we might end up underestimating the amount of confounding learned by the algorithm. In practice, one should experiment with distinct discretizations.


Building upon this test, we develop two novel statistical tools to deal with confounding. First, by noticing that the location of the restricted permutation null distribution provides a natural measure of the amount of confounding signal learned by the algorithm, we adopt the average of the restricted permutation null as a test statistic, and develop a novel statistical test to detect confounding learning per se. Second, motivated by the fact that restricted permutations provide an alternative approach to compute partial correlations (a result that we prove in Theorem 1), we develop a novel approach to correct for the influence of the confounders. The basic idea is to estimate the ``corrected predictive performance" of a learner, by removing the contribution of the confounders from the observed predictive performance using a mapping between the restricted and standard permutation null distributions\footnote{Here, we denote as ``standard permutations" the usual permutation scheme where the unrestricted shuffling of the response destroys the association between the response and features and between response and confounders.}.

A key practical application of our tools is to evaluate the extent to which any adjustment method has succeeded in removing the influence of the confounders. This is important in practice since most methods that aim to adjust for confounders rely on assumptions, and it is generally unclear how robust these methods are to violations of these assumptions. (For instance, the ``adjusted features" approach assumes linear associations between confounders and features, and might be ineffective, or even prone to artifacts, when these associations are mostly non-linear. The weights employed in the inverse probability weighting approaches need to be estimated from the data, and often depend on strong modeling assumptions.)

\section{The permutation approach}

First, we introduce some notation. Throughout the text we let $m$ represent a predictive performance metric; $\bfX$ the feature data matrix; $\bfy$ the response data vector; and $\bfc$ the observed confounder data vector\footnote{Note that if multiple confounders are available, than $\bfc$ can represent the combined categorical confounder vector, generated by pasting together the multiple confounder vectors. For instance, if our confounders are a discretized age variable with levels ``young", ``middle age", and ``senior", and a gender variable with levels ``male" and ``female", then the combined confounder has the following six levels: ``young male", ``young female", ``middle age male", ``middle age female", ``senior male", and ``senior female".}. We reserve $F$ to represent a cumulative distribution function (c.d.f.) of an arbitrary random variable, and $\Phi$ to represent the c.d.f. of a standard normal variable. We let $F_{\pi^\ast}$ and $F_{\pi^{\ast\ast}}$ represent, respectively, the restricted and standard permutation null distributions (which we describe in the next subsections), and $\hat{F}_{\hat{\pi}^\ast}$ and $\hat{F}_{\hat{\pi}^{\ast\ast}}$ represent the respective Monte Carlo versions of these permutation distributions. We let $Y^\ast$ and $Y^{\ast\ast}$ represent restricted and standard permutations of the response variable $Y$, and $m^\ast$ and $m^{\ast\ast}$ represent the predictive performance metrics computed with the respective permuted responses.

\subsection{Detecting response learning in the presence of confounding}

Restricted permutations represent an standard approach to account for the influence of confounders in randomization tests (Good 2000). They have also been used in the context of predictive modeling (Rao et al. 2017). Here, we describe the approach in detail, as it provides the necessary background for the novel contributions described in the next sections.

\begin{figure}[!h]
  \begin{center}
  \centerline{\includegraphics[width=\linewidth, bb = 0 100 1400 560]{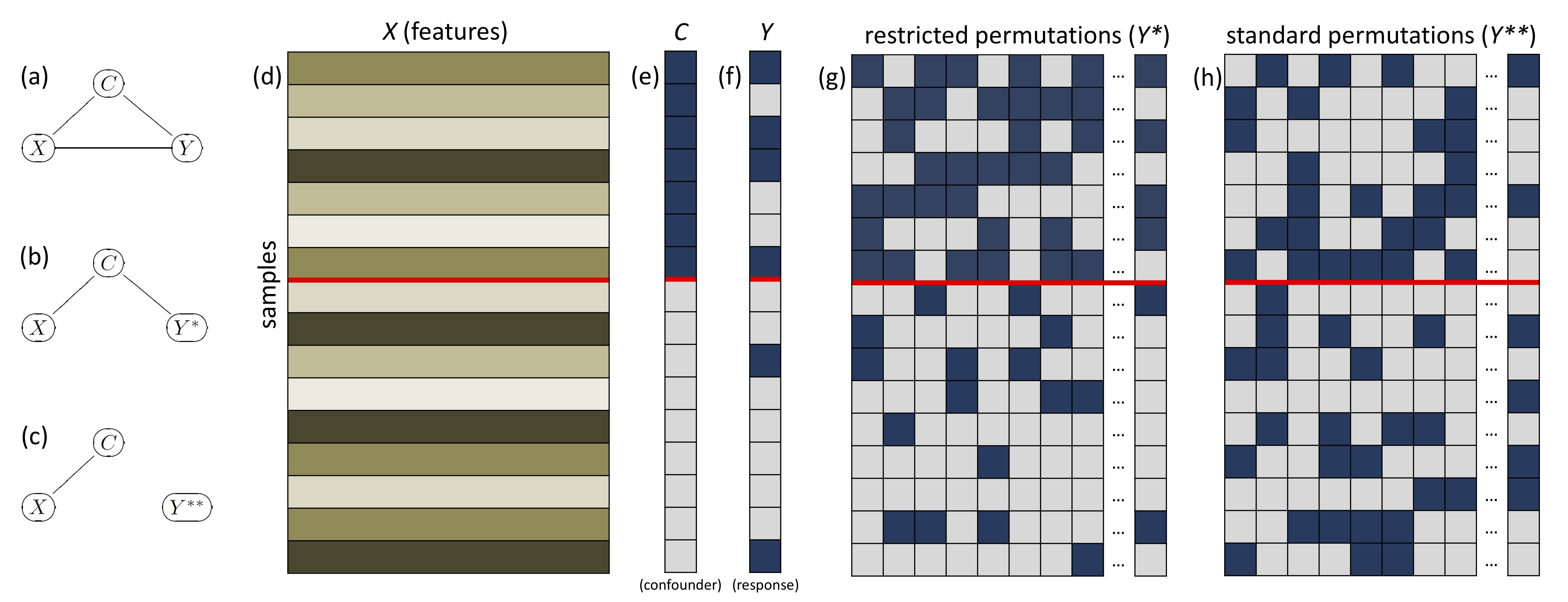}}
  \end{center}
  \caption{Panel a shows an undirected dependency graph (UDG) representing the associations between the features, $X$, the response, $Y$, and the observed confounders, $C$. (In an UDG, an undirected edge between two nodes means that the variables are associated even when we condition on the remaining variables.) Panels d, e, and f, represent the feature, confounder, and response data, respectively. In this cartoon example, we have 16 samples, and both $C$ and $Y$ are binary (light and dark cells represent 0 and 1 values, respectively). The confounder vector (panel e) was sorted, and the red line splits the data relative to the levels of $C$ (i.e., the top 7 samples have confounding value 1, while the bottom 9 have confounding value 0). Note that in panel f we have 4 positive response values (dark cells) above the red line, and 2 below it. Panel g illustrates the restricted permutation scheme. Each column shows a distinct permutation. In all permutations, we still have 4 dark cells above the red line and 2 below it. Panel b shows the UDG representation of these relations (note that conditional on the confounder the features and the restricted shuffled response, $Y^\ast$, become independent). Panel h illustrates the standard permutation scheme, where we shuffle the response values freely across the entire response vector (now, each column is no longer constrained to have 4 dark cells above the red line and 2 below it). The standard permutations destroy the association between $Y$ and $C$ and between $Y$ and $X$. Panel c shows the UDG representation in this case.}
  \label{fig:stratifiedperm}
\end{figure}

When the relationship between the features, $\bfX$, and the response, $\bfy$, is influenced by an observed set of confounders, $\bfc$, the total association between $\bfX$ and $\bfy$ can be partitioned into a component due to the direct association between $\bfX$ and $\bfy$ (due to a potential causal relation between $\bfX$ and $\bfy$, or to the presence of additional unmeasured confounders associated with $\bfX$ and $\bfy$), and into an indirect component where part of the association between features and response is explained by the association between $\bfX$ and $\bfc$, and the association between $\bfc$ and $\bfy$. Figure \ref{fig:stratifiedperm}a provides a graphical model representation of these associations.

In order to evaluate whether a machine learning algorithm has learned about the response variable (even when confounding is present) we need to generate a permutation null distribution where the direct association between the response and the features is destroyed, while the indirect association mediated by the confounder is still preserved (Figure \ref{fig:stratifiedperm}b). To this end, we generate a restricted permutation null distribution, as described in Algorithm \ref{alg:restrictedShuffling}, by separately shuffling the response data within each level of the confounder (as illustrated in Figure \ref{fig:stratifiedperm}e-g).

\begin{algorithm}
\caption{Restricted Monte Carlo permutation null distribution for performance metric $m$}\label{alg:restrictedShuffling}
\begin{algorithmic}[1]
\State \textbf{Input}: Number of permutations, $b$; $\bfX$; $\bfy$; $\bfc$; training and test set indexes, $i_{train}$, $i_{test}$
\State Split $\bfX$, $\bfy$ and $\bfc$ into training and test sets
\For{$i = 1, 2, \ldots, b$}
  \State $\bfy^{\ast}_{train} \leftarrow \mbox{RestrictedShuffle}(\bfy_{train}, \bfc_{train})$, and $\bfy^{\ast}_{test} \leftarrow \mbox{RestrictedShuffle}(\bfy_{test}, \bfc_{test})$
  \State Train a machine learning algorithm on the $\bfX_{train}$ and $\bfy_{train}^{\ast}$ data
  \State Evaluate the algorithm on the $\bfX_{test}$ and $\bfy_{test}^{\ast}$ data
  \State Record the value of the performance metric, $m^\ast_i$, on the shuffled data
\EndFor
\State \textbf{Output}: $m^\ast_1$, $m^\ast_2$, \ldots, $m^\ast_b$
\end{algorithmic}
\end{algorithm}

\begin{figure}[!h]
  \centering
  \centerline{\includegraphics[width=\linewidth, bb = 0 0 790 220]{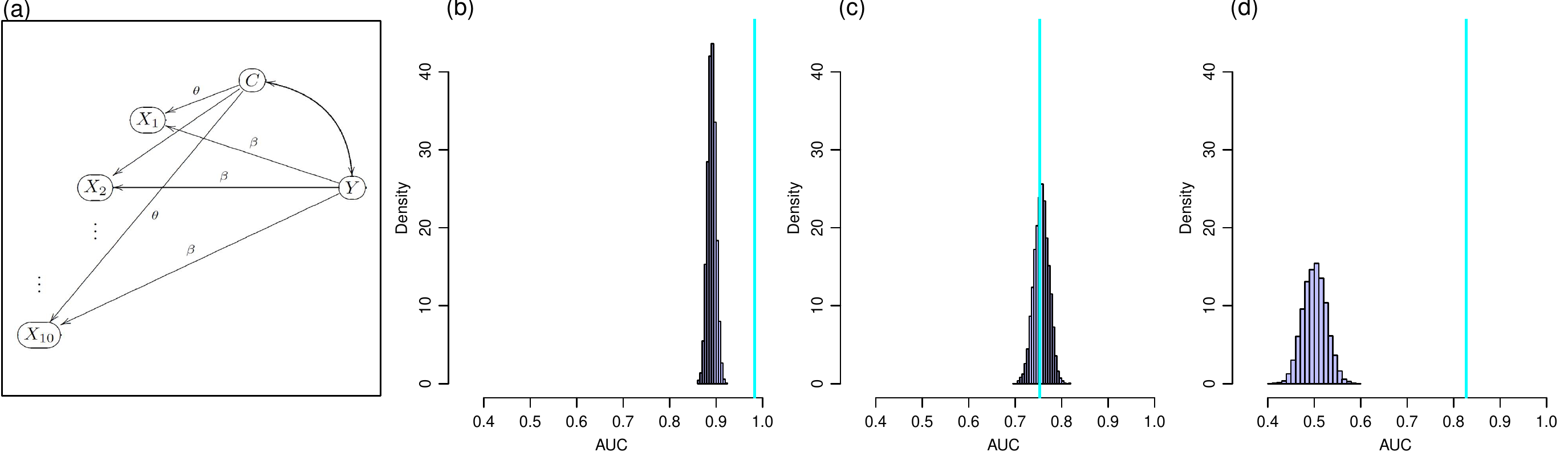}}
  \caption{This figure illustrates the generation of the restricted Monte Carlo permutation null for the AUC metric using synthetic data. Panel a shows the causal graph representing the model used to simulate the data. (Details about this model are provided in Section 3.1.) Panels b-d show the restricted Monte Carlo permutation null distributions (blue histograms) for data generated with distinct strengths of confounding. The cyan lines represent the observed AUC values. In panel b, we simulated a very strong association between $C$ and $Y$ ($\hat{cor}(C, Y) = 0.8$), and causal influences of $C$ on $\bfX$ ($\theta = 1$), and of $Y$ on $\bfX$ ($\beta = 1$). The restricted null is located close to 0.9, far away from the baseline random guess value (0.5 for the AUC metric), but still below the observed AUC value (cyan line around 0.98). This shows that while a lot of the predictive performance of the classifier is explained by the confounder, the direct association between $Y$ and $\bfX$ also contributes to it. Hence, in this example, the classifier was still able to learn the disease signal, in spite of the strong confounding influence. In this example, the permutation p-value for testing $H_0^\ast$ vs $H_1^\ast$ is smaller than $9.9 \times 10^{-5}$ (we used 10,000 permutations to generate the null). In panel c, we simulated moderate/strong association between $C$ and $Y$ ($\hat{cor}(C, Y) = 0.6$) and a causal influence of $C$ on $\bfX$ ($\theta = 1$), but not of $Y$ on $\bfX$ ($\beta = 0$). In this case while the restricted null is, again, located far way from 0.5, the observed AUC score (cyan line around 0.75) is located inside the range of the permutation null showing that the classifier is detecting the confounding signal, but not the response signal (as expected since we set $\beta = 0$ in this example). In this example, the permutation p-value is 0.59. Finally, in panel d we simulated data with causal influences of $C$ on $\bfX$ ($\theta = 1$) and $Y$ on $\bfX$ ($\beta = 1$), but no association between $C$ and $Y$ (that is, we generated unconfounded data). As, expected, the restricted permutation distribution is centered around 0.5, while the observed AUC is above 0.8. These examples illustrate that the restricted permutation null is centered away from the baseline random guess value whenever confounding is present (and that this shift can be used to informally infer the presence of confounding).}
  \label{fig:strataucillustrations}
\end{figure}

Figure \ref{fig:strataucillustrations} show examples of the restricted permutation null distributions (for the AUC metric) generated with varying amounts of confounding. These examples show that the restricted permutation null is always centered away from the baseline random guess value whenever confounding is present, and illustrate that this shift can be used to informally infer the presence of confounding.

The restricted Monte Carlo permutation null allows us to test the hypotheses,
\begin{align*}
H_0^\ast &: \mbox{the algorithm is not learning the response signal,} \\
H_1^\ast &: \mbox{the algorithm is learning the response signal.}
\end{align*}

A permutation p-value for testing $H_0^\ast$ is computed as the proportion of times that the performance metric computed in shuffled response data was equal to or better than the performance metric computed with the original (un-shuffled) response vector, $m_o$.

\subsection{Correcting for the influence of confounders in machine learning predictions}

Let $E_{\pi^\ast}[M^\ast]$ represent the expectation of the restricted permutation null distribution. This quantity represents a natural candidate to measure the contribution of confounding to the learner's predictive ability, as it measures the algorithm's performance after the algorithm's ability to learn the direct association between the response and the features has been neutralized by the restricted shuffling of the response data. The observed metric value, $m_o$, on the other hand, captures the contributions of both response learning and confounder learning to the predictive performance. It is reasonable to expect that the contribution of the response alone should be a function of the difference between $m_o$ and $E_{\pi^\ast}[M^\ast]$, and we would intuitively expect that the corrected performance metric, $m_c$, to assume the form,
\begin{equation}
m_c \, = \, f\big(m_o \, - \, E_{\pi^\ast}[M^\ast]\big)~.
\label{eq:intuitivecorrection}
\end{equation}

As a matter of fact, the following result (proved in the Appendix) formalizes this intuition for linear measures of statistical association.
\begin{theorem}
Let $C$ represent a categorical variable, $X$ and $Y$ represent arbitrary random variables, and $Y^\ast$ represent a restricted permutation of $Y$ relative to the levels of $C$. Let $Cov(X, Y)$ represent the covariance between $X$ and $Y$, and $pCov(X, Y \mid C)$ represent the partial covariance between $X$ and $Y$ given $C$. An alternative formula for the computation of $pCov(X, Y \mid C)$, based on the restricted permutations of $Y$ with respect to the levels of $C$, is given by,
$$
pCov(X, Y \mid C) \, = \, Cov(X, Y)  \, - \, E_{\pi^\ast}\left[ Cov(X, Y^\ast) \right]~.
$$
Similarly, the partial correlation $pCor(X, Y \mid C)$ can be expressed as,
$$
pCor(X, Y \mid C) \, = \, \left(\frac{Var(X) \, Var(Y)}{Var(X \mid C) \, Var(Y \mid C)}\right)^{\frac{1}{2}} \, \times \, \big(Cor(X, Y) - E_{\pi^\ast}\left[ Cor(X, Y^\ast) \right]\big)~.
$$
\end{theorem}

Note that, for the covariance metric the function $f()$ in equation (\ref{eq:intuitivecorrection}) corresponds to the identity function, while for the correlation metric it simply re-scales the difference $m_o \, - \, E_{\pi^\ast}[M^\ast]$.

Theorem 1 shows that the total amount of linear association between two arbitrary random variables measured by $Cov(X, Y)$ can be partitioned into the $E_{\pi^\ast}\left[ Cov(X, Y^\ast) \right]$ component, that measures the amount of association between $X$ and $Y$ that is due exclusively to the confounder $C$ (i.e., the indirect association), and into the partial covariance component\footnote{Note that in Theorem 1 we adopt the partial covariance metric to represent the amount of association between $X$ and $Y$ that remains after we remove the indirect association due to $C$, instead of the conditional covariance metric (which, one might argue, might represent another natural metric in this context). Observe, however, that Theorem 1 assumes linear associations, and under this assumption the partial and conditional covariance measures are closely related. (Theorem 1 of Baba, Shibata, and Sibuya (2004) shows that, in this setting, the partial covariance equals the expected conditional covariance, while Corollary 2 proves the equality of these quantities when the conditional variance/covariance matrix is free of the conditioning variable, as is the case for the multivariate normal distribution).}, $pCov(X, Y \mid C)$, that measures the remaining amount of association between $X$ and $Y$ that is not explained by the confounder $C$ (i.e., the direct association).

We point out, however, that the above result is not exclusively available for linear measures of association. In the Appendix, we also present and prove a similar result (Theorem 2) involving the distance correlation (Szekely, Rizzo, and Bakirov 2007) and partial distance correlation (Szekely and Rizzo 2014), showing that the restricted permutations can also be used to decompose non-linear measures of statistical association.

Motivated by these results, we now propose a general approach to compute the corrected performance metric, $m_c$, for an arbitrary metric $m$. It requires, however, the computation of a second permutation distribution, $F_{\pi^{\ast\ast}}$, denoted the ``standard" permutation null, that is generated in the usual way by freely shuffling the response values across the entire response vector (Figure \ref{fig:stratifiedperm}h). Note that the standard shuffling destroys the association between the response and the features, as well as, the association between the response and the confounder (Figure \ref{fig:stratifiedperm}c), and can be used to test the hypotheses,
\begin{align*}
H_0^{\ast\ast} &: \mbox{the algorithm is not learning the response and confounding signals,} \\
H_1^{\ast\ast} &: \mbox{the algorithm is learning the response and/or the confounding signal.}
\end{align*}
A Monte Carlo estimate of the standard permutation null is generated as described in Algorithm 1, except that the restricted shuffling of the response data in step 4 is replaced by standard shuffling.

\subsubsection{The confounding corrected metric}

As pointed before, the observed metric $m_o$ captures the contributions of both response and confounder learning. In order to estimate the confounding corrected value $m_c$ we need to determine what value would the observed performance metric have assumed, had the response variable not been associated with the confounder. In other words we need to map a value sampled from a distribution where the response and confounder are associated to a distribution where they are not.

To this end, we construct a mapping from the restricted permutation null distribution (where the association between the response and the confounder is preserved) to the standard permutation null (where this association is removed). An obvious mapping would be to define $m_c = m_o - a_{\hat{\pi}^{\ast}} + a_{\hat{\pi}^{\ast\ast}}$, where $a_{\hat{\pi}^{\ast}}$ and $a_{\hat{\pi}^{\ast\ast}}$ correspond, respectively, to the sample mean of $\hat{F}_{\hat{\pi}^\ast}$ and $\hat{F}_{\hat{\pi}^{\ast\ast}}$. This mapping, however, only focus on the means and fails to take into consideration the different spreads of the restricted and standard permutation null distributions. Ideally, we should define a mapping that accounts for the entire probability distributions. Therefore, we define and estimate the corrected metric $m_c$ by equating $F_{\pi^{\ast\ast}}(m_c)$ to $F_{\pi^\ast}(m_o)$,
\begin{equation}
F_{\pi^{\ast\ast}}(\hat{m}_c) \, = \, F_{\pi^\ast}(m_o) \; \Leftrightarrow \; \hat{m}_c \, = \, F_{\pi^{\ast\ast}}^{-1}(F_{\pi^\ast}(m_o))~.
\label{eq:correctedmetricperm}
\end{equation}

Note that equating $F_{\pi^\ast}(m_o)$ to $F_{\pi^{\ast\ast}}(m_c)$  is equivalent to equating the p-value for testing $H_0^\ast$ vs $H_1^\ast$ to the p-value for testing $H_0^{\ast\ast}$ vs $H_1^{\ast\ast}$. (For metrics where smaller values indicate better predictive performance (e.g., mean squared error) we have that $F_{\pi^\ast}(m_o)$ corresponds to the p-value for testing $H_0^\ast$. For metrics where larger values indicate better performance (e.g, AUC), the p-value is given by $1 - F_{\pi^\ast}(m_o)$, and we get the same correction formula by matching this p-value to $1 - F_{\pi^{\ast\ast}}(m_c)$.)

\begin{figure}[!h]
  \centering
  \centerline{\includegraphics[width=5in]{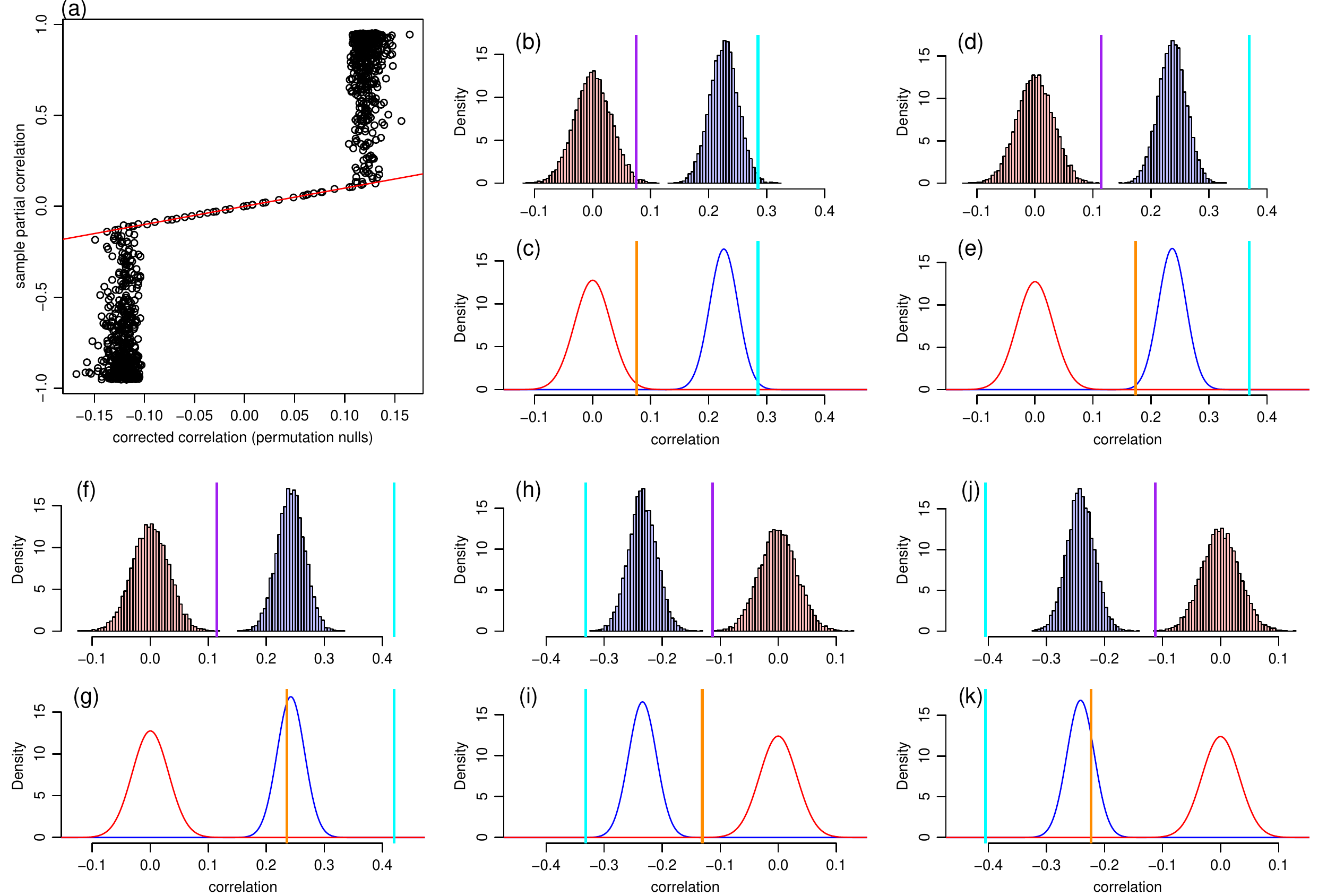}}
  \caption{{Comparison of partial correlation and corrected correlation values across 1,000 simulated data sets (generated as described in Section 6). Panel a compares the corrected correlation value (x-axis), computed with the formula $\hat{m}_c = \hat{F}_{\hat{\pi}^{\ast\ast}}^{-1}(\hat{F}_{\hat{\pi}^\ast}(m_o))$, against the partial correlation estimated using the standard partial correlation formula (y-axis). The red line represents the identity line, where the corrected correlation score exactly matches the partial correlation score. Note that for some of the simulated data sets the two formulas provide very close scores, while, for most of the simulations, the corrected correlation tends to be truncated around -0.12 and 0.12. Panels b-k explain why. In all panels, the blue and red histograms represent, respectively, the restricted and standard permutation null distribution, while the blue and red bell shaped curves represent the respective normal approximations. The cyan line represents the observed correlation score. The purple line represents the corrected correlation, computed using $\hat{m}_c = \hat{F}_{\hat{\pi}^{\ast\ast}}^{-1}(\hat{F}_{\hat{\pi}^\ast}(m_o))$ (and from now on denoted as the ``purple" correction), while the orange line represents correlations corrected according to the formula $\hat{m}_c = (m_o - a_{\hat{\pi}^{\ast}}) s_{\hat{\pi}^{\ast\ast}}/s_{\hat{\pi}^{\ast}} + a_{\hat{\pi}^{\ast\ast}}$ (denoted ``orange" correction). Points falling at the red line in panel a represent simulations where the observed correlation score falls inside the range of the restricted permutation null, as illustrated in panels b and c. (Note how the purple line in panel b matches the orange line in panel c). Panel d shows the case where the observed correlation falls above the range of the restricted permutation null. In this case, the corrected value (purple line) corresponds to the upper boundary of the red histogram since $\hat{F}_{\hat{\pi}^\ast}(m_o) = 1$ for any observed correlation value above the range of the blue histogram and $\hat{F}_{\hat{\pi}^{\ast\ast}}^{-1}(1)$ corresponds to the upper boundary of the red histogram. Note, however, that the corrected value based on the asymptotic approximation (orange line in panel e) is higher than the purple line. Panel f shows an example where the observed correlation is even larger, but the ``purple" corrected value is still similar to the previous example (compare the purple lines in panels d and f). The ``orange" corrected value, on the other hand, is larger (compare the orange lines in panels e and g). The examples in panels b-g illustrate why the corrected values computed with $\hat{m}_c = \hat{F}_{\hat{\pi}^{\ast\ast}}^{-1}(\hat{F}_{\hat{\pi}^\ast}(m_o))$ get truncated around 0.12 in panel a. Similarly, panels h-k illustrate that, when the observed correlation is negative, the ``purple" corrected values correspond to the minimum of the standard permutation null distributions (since $\hat{F}_{\hat{\pi}^\ast}(m_o) = 0$ and $\hat{F}_{\hat{\pi}^{\ast\ast}}^{-1}(0)$ corresponds to the lower boundary of the standard permutation null). Hence, we also observe a truncation of the ``purple" corrected correlations around -0.12 in panel a. Finally, note that the asymptotic approximation fixes this issue, so that the ``orange" corrected correlations closely approximate the partial correlations.}}
  \label{fig:quantileissue}
\end{figure}

In general, it is unfeasible to simulate $F_{\pi^\ast}$ and $F_{\pi^{\ast\ast}}$ exactly due the large number of possible permutations. Presumably, one could rely on Monte Carlo estimates of these distributions and estimate $m_c$ as $\hat{m}_c = \hat{F}_{\hat{\pi}^{\ast\ast}}^{-1}(\hat{F}_{\hat{\pi}^\ast}(m_o))$, where $\hat{F}_{\hat{\pi}^\ast}(x)$ represents the $x$th sample percentile of the Monte Carlo restricted null, and $\hat{F}_{\hat{\pi}^{\ast\ast}}^{-1}(x)$ represents the $x$th sample quantile of the Monte Carlo standard null. Observe, however, that this approach only works when $m_o$ falls inside the range of $\hat{F}_{\hat{\pi}^\ast}$. As illustrated in Figure \ref{fig:quantileissue}, the corrected metric value gets artificially ``truncated" whenever the observed correlation is outside the range of the restricted permutation null. Therefore, in practice, we need to use analytical approximations for the restricted and standard permutation null distributions in order to estimate $m_c$.

\begin{figure}[h]
  \centering
  \centerline{\includegraphics[width=3in, bb = 0 0 1090 780]{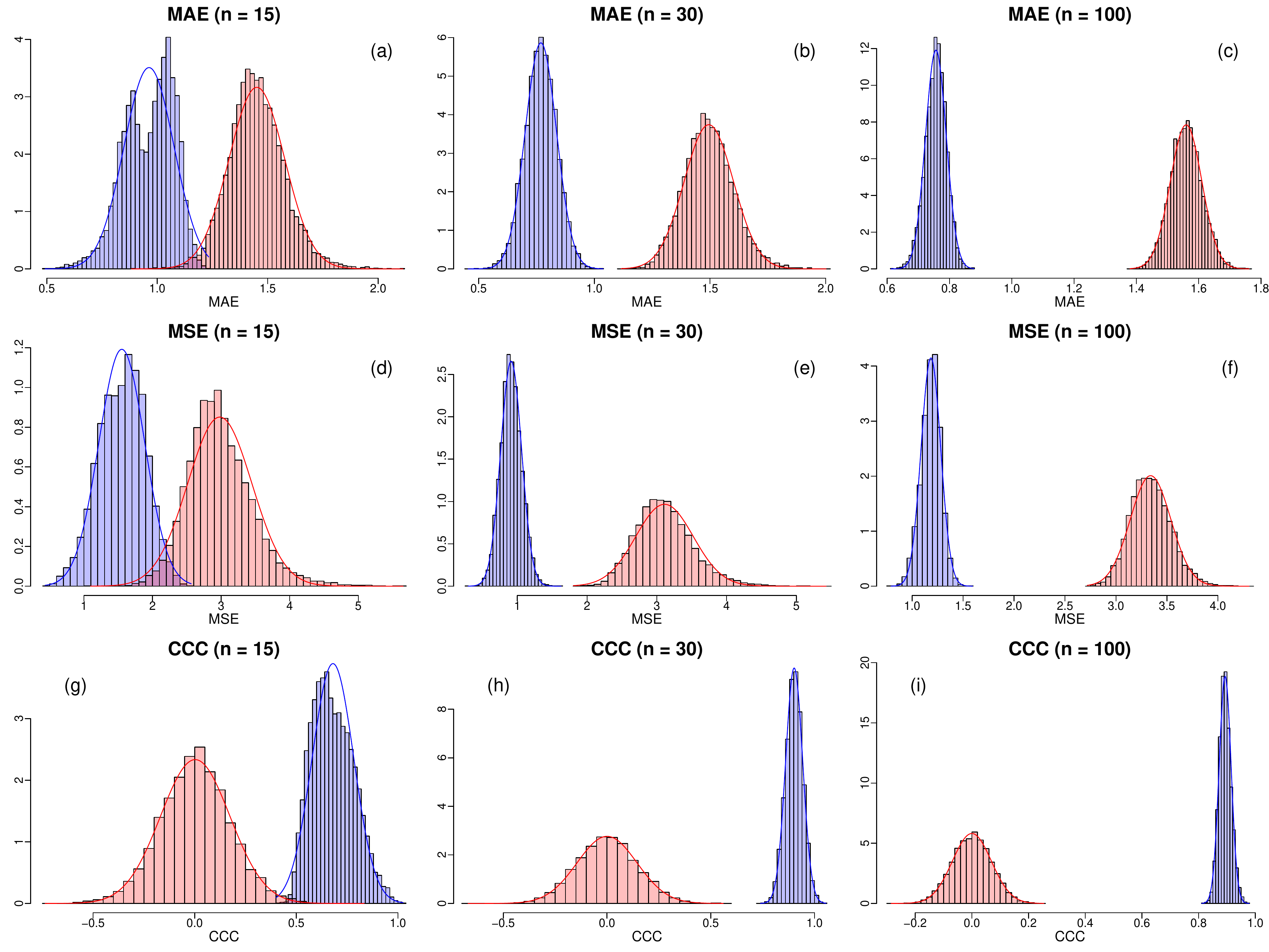}}
  \caption{Asymptotic normality of common performance metrics for regression problems. Panel a, b and c show the restricted (blue histograms) and standard (red histograms) permutation null distributions for the mean absolute error (MAE) metric, generated using training and test sets of size 15, 30, and 100, respectively. Panels d, e, and f, show similar plots for the mean squared error (MSE) metric, while panels g, h, and i, show the same plots for the concordance correlation coefficient (CCC) metric. The data was simulated as described in the Appendix, except that we adopted exponential instead of gaussian errors for the generation of the response. For all metrics we observe a poor approximation for test sets with 15 samples, but reasonable approximations for test sets with 30 or 100 samples.}
  \label{fig:regrasymptotic}
\end{figure}

\begin{figure}[h]
  \centering
  \centerline{\includegraphics[width=3in, bb = 0 0 1090 530]{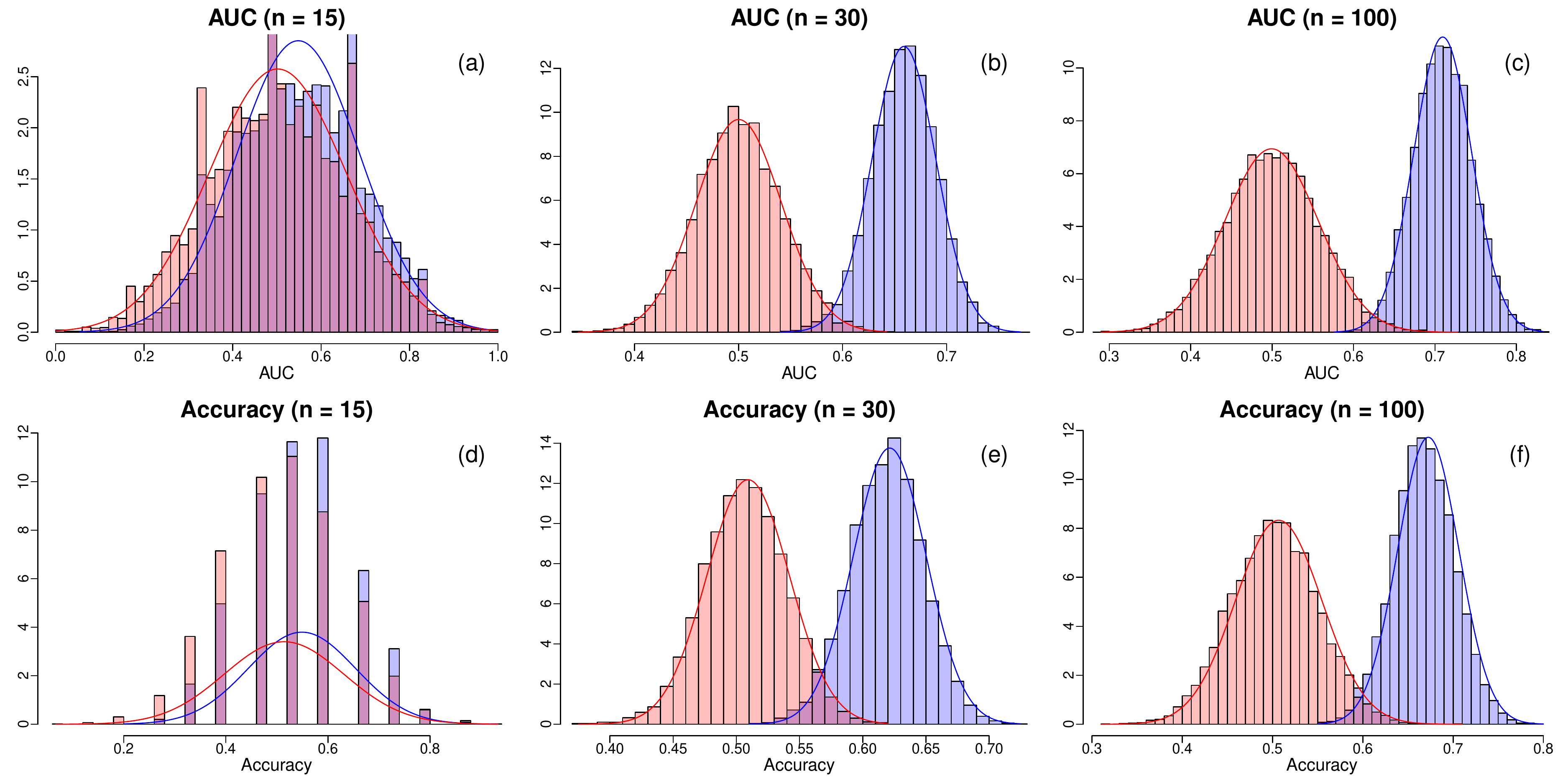}}
  \caption{Asymptotic normality of common performance metrics for classification problems. Panel a, b and c show the restricted (blue histograms) and standard (red histograms) permutation null distributions for the AUC metric, generated with training and test sets of size 15, 30, and 100, respectively. Panels d, e, and f, show similar plots for the accuracy metric. The data was simulated as described in Section 5. For all metrics we observe a poor approximation for test sets with 15 samples (note the discreteness of the permutation distributions), but reasonable approximations for test sets with 30 or 100 samples.}
  \label{fig:classasymptotic}
\end{figure}

Fortunately, because popular performance metrics such as the mean square error, mean absolute error, and the classification accuracy correspond to averages, while metrics such as the AUC correspond to a generalized U-statistics (DeLong, DeLong and Clarke-Pearson 1988; Lehmann 1951), we have that the distribution of these statistics can be well approximated by Gaussian distributions when the test set is large enough (due to central limit theorems associated with averages, and to the asymptotic normality of (generalized) U-statistics (Hoeffding 1948, Serfling 1980)). Figures \ref{fig:regrasymptotic} and \ref{fig:classasymptotic} provide a few illustrative examples. Hence, in practice, we will often be able to approximate $\hat{F}_{\hat{\pi}^\ast}$ and $\hat{F}_{\hat{\pi}^{\ast\ast}}$ by,
\begin{equation}
\hat{F}_{\hat{\pi}^{\ast}} \, \approx \, N(a_{\hat{\pi}^{\ast}} \, , \, s^2_{\hat{\pi}^{\ast}})~, \hspace{0.5cm} \hat{F}_{\hat{\pi}^{\ast\ast}} \, \approx \, N(a_{\hat{\pi}^{\ast\ast}} \, , \, s^2_{\hat{\pi}^{\ast\ast}})~,
\label{eq:approxgaussian}
\end{equation}
where $s^2_{\hat{\pi}^{\ast}}$ and $s^2_{\hat{\pi}^{\ast\ast}}$ correspond, respectively, to the sample variances of $\hat{F}_{\hat{\pi}^\ast}$ and $\hat{F}_{\hat{\pi}^{\ast\ast}}$, and $a_{\hat{\pi}^{\ast}}$ and $a_{\hat{\pi}^{\ast\ast}}$ represent, as before, the respective sample means. Now, by replacing $F_{\pi^\ast}$ and $F_{\pi^{\ast\ast}}$ in equation (\ref{eq:correctedmetricperm}) by the approximate Gaussian distributions in (\ref{eq:approxgaussian}) we have that,
\begin{align}
\hat{F}_{\hat{\pi}^{\ast\ast}}(\hat{m}_c) \, &\approx \, \Phi\big((\hat{m}_c - a_{\hat{\pi}^{\ast\ast}})/s_{\hat{\pi}^{\ast\ast}}\big) \, \\ \nonumber
&= \, \Phi\big((m_o - a_{\hat{\pi}^{\ast}})/s_{\hat{\pi}^{\ast}}\big) \, \approx \, \hat{F}_{\hat{\pi}^{\ast}}(m_o)~,
\end{align}
and we can estimate $\hat{m}_c$ by,
\begin{equation}
\hat{m}_c \, = \, (m_o - a_{\hat{\pi}^{\ast}}) \, \frac{s_{\hat{\pi}^{\ast\ast}}}{s_{\hat{\pi}^{\ast}}} + a_{\hat{\pi}^{\ast\ast}}~.
\label{eq:correctedmetric}
\end{equation}

Note that the above correction formula still have the same format as equation (\ref{eq:intuitivecorrection}), with $f()$ performing a re-scaling and translation of the difference, $m_o - \hat{E}_{\hat{\pi}^\ast}[M^\ast]$. Furthermore, for the correlation metric, $a_{\hat{\pi}^{\ast\ast}} = \hat{E}_{\hat{\pi}^{\ast\ast}}[M^{\ast\ast}] \approx 0$, and the correction has similar format as the formula in Theorem 1.

\begin{figure}[!h]
  \centering
  \centerline{\includegraphics[width=5in, bb = 0 10 790 200]{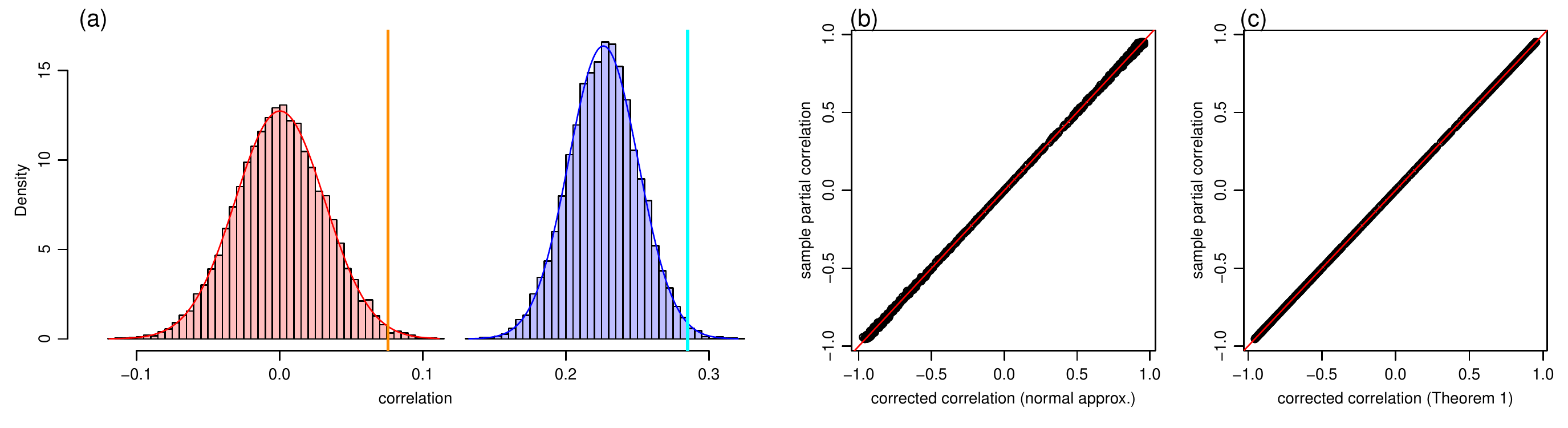}}
  \caption{Panel a shows an example of the restricted (blue) and standard (red) Monte Carlo permutation null distributions for the correlation metric (see the Appendix for the simulation details). The cyan line shows the observed correlation, while the orange one shows the corrected correlation value (from equation \ref{eq:correctedmetric}). The blue and red curves correspond to the normal approximations in equation (\ref{eq:approxgaussian}). Note that the tail probabilities to the right of the cyan and orange lines are the same (i.e., the p-values are preserved). Panels b and c compare our correction formulas against the sample partial correlation estimates.}
  \label{fig:corrillustrations}
\end{figure}

Figure \ref{fig:corrillustrations}a illustrates the application of formula \ref{eq:correctedmetric} to the correlation metric. Figure \ref{fig:corrillustrations}b illustrates that the correction formula (x-axis) was able to recapitulate very well the sample partial correlation values (y-axis), computed as,
\begin{equation}
\hat{pcor}(X, Y \mid C) = \frac{\hat{cor}(X, Y) - \hat{cor}(X, C) \, \hat{cor}(Y, C)}{[(1 - \hat{cor}(X, C)^2)(1 - \hat{cor}(Y, C)^2)]^{1/2}}~.
\end{equation}
For completeness, Figure \ref{fig:corrillustrations}c compares the partial correlation values computed with the sample version of the formula presented in Theorem 1, against the sample partial correlation values.

\subsection{A statistical test to detect confounding}

As described before, the presence of confounding will shift the restricted permutation null distribution away from the baseline random guess value, and this shift can be used to informally infer the presence of confounding. Here, we present a hypothesis test to formally test the hypotheses,
\begin{align*}
H_0^c &: \mbox{the machine learning algorithm has not learned the confounding signal} \\
H_1^c &: \mbox{the machine learning algorithm has learned the confounding signal.}
\end{align*}
We adopt the sample mean of the restricted permutation null,
\begin{equation}
\bar{M}^\ast = \frac{1}{b} \sum_{i = 1}^{b} M_i^\ast~,
\end{equation}
as a test statistic, since it represents a natural measure of confounding. Note that under the null hypothesis that an algorithm has not learned the confounding signal, the restricted permutation null will have the same distribution as the standard permutation null. Hence, for large enough test sets we have that $M^\ast \approx N(a_{\hat{\pi}^{\ast\ast}} \, , \, s^2_{\hat{\pi}^{\ast\ast}})$, and our test statistic is asymptotically distributed as,
\begin{equation}
\bar{M}^\ast \, \approx \, N\left(a_{\hat{\pi}^{\ast\ast}} \, , \, \frac{s^2_{\hat{\pi}^{\ast\ast}}}{b}\right)~.
\end{equation}
Note that the variance of this null distribution depends on the number of permutations ($b$) used to generate the restricted permutation null, and gets smaller as we increase $b$. As a consequence, we can easily obtain a statistically significant result by increasing the number of permutations. For example, suppose we are working on a classification problem using the AUC metric, and the sample average of the restricted permutation null, $\bar{auc}^\ast$, is given by 0.51. If the confounding effect is small but real, we can improve the test's statistical significant by increasing the number of permutations, since our test statistic estimate will converge to a small but larger than 0.5 value, while the confounding null distribution will get more and more peaked around 0.5 (the baseline random guess value for the AUC metric).

In order to avoid this artifact, we restrict $b$ to be equal to the size of the test set. By doing so, we guarantee that we will only be able to detect small confounding effects when we are truly well powered to do so. In Section 3, we report the results of a simulation study evaluating the empirical performance of the confounding test (Figure 3b), where we set the number of permutations to be equal to the test set sample size. We observed good power to detect confounding under $H_1^c$, and well controlled type I error rates under $H_0^c$.

In the Appendix, we also present a general permutation scheme (Algorithm 2) that does not require asymptotic normality assumption, and can be used with small test sets.

\subsection{Analytical results for the AUC metric}

It has been shown\cite{bamber1975} that, when there are no ties in the predicted class probabilities used for the computation of the $AUC$, the test statistic of the Wilcoxon rank sum test (also known as the Mann-Whitney U test), $U$, is related to the $AUC$ statistic by, $U = n_n\,n_p (1 - AUC)$, where $n_n$ and $n_p$ represent the number of negative and positive labels in the test set (see Section 2 of reference\cite{mason2002} for details).

For large test sets, and under the null hypothesis $H_0^{\ast\ast}$ (i.e., that the machine learning algorithm has not learned the response and the confounding signal), this distribution can be approximated\cite{mason2002} by
\begin{equation}
U \, \approx \, N\left( \frac{n_n\,n_p}{2} \; , \; \frac{n_n\,n_p (n_n + n_p + 1)}{12} \right)~.\footnote{In the presence of ties, a slightgly better approximation, is given by,
$$
U \, \approx \, N\left( \frac{n_n\,n_p}{2} \; , \; \frac{n_n\,n_p (n_n + n_p + 1)}{12} - \frac{n_n\,n_p}{12 \, n \, (n - 1)} \sum_{k=1}^{\tau} t_k(t_k-1)(t_k+1) \right)~,
$$
where $\tau$ is the number of groups of ties, and $t_k$ is the number of ties in group $k$\cite{mason2002}. We have compared both approximations and did not detect any discernible differences in the results. For this reason, we adopted the simpler approximation in equation (\ref{eq:uapprox}).}
\label{eq:uapprox}
\end{equation}
Now, from the relation $AUC = 1 - U/(n_n\,n_p)$ it follows that,
\begin{equation}
AUC \, \approx \, N\left(\frac{1}{2} \, , \, \frac{n_n + n_p + 1}{12 \, n_n \, n_p}\right)~,
\end{equation}
under the null $H_0^{\ast\ast}$, so that $F_{\pi^{\ast\ast}}$ can be approximated by the above normal distribution.

Using the ``p-value matching" criterion, $F_{\pi^{\ast\ast}}(auc_c) = F_{\pi^\ast}(auc_o)$, it follows that,
\begin{equation}
\Phi\left( \frac{auc_c - 0.5}{\sigma} \right) \, \approx \, F_{\pi^\ast}(auc_o) \;\;\; \Leftrightarrow \;\;\; auc_c \, \approx \, \Phi^{-1}(F_{\pi^\ast}(auc_o)) \, \sigma \, + \, 0.5~,
\end{equation}
where $\Phi$ represents the cumulative density function of a standard normal random variable, and $\sigma = \sqrt{(n_n + n_p + 1)/(12 \, n_n \, n_p)}$.

Now, observe that because the $AUC$ is a generalized U-statistic \cite{lehmann1951,serfling1980} it will also be asymptotically distributed as a normal random variable (even under the alternative). Hence, for large sample sizes,
\begin{equation}
F_{\pi^\ast} \, \approx \, N\left(a_{\hat{\pi}^\ast} \, , \, s_{\hat{\pi}^\ast}^2 \right)~, \hspace{0.5cm} a_{\hat{\pi}^\ast} = \hat{E}_{\hat{\pi}^\ast}[AUC^\ast]~, \hspace{0.5cm} s_{\hat{\pi}^\ast}^2 \, = \, \hat{Var}_{\hat{\pi}^\ast}(AUC^\ast)~,
\end{equation}
where $\hat{E}_{\hat{\pi}^\ast}$ and $\hat{Var}_{\hat{\pi}^\ast}$ represent the sample average and sample variance of the restricted permutation null distribution of $AUC$ scores. The confounder corrected $AUC$ score is then estimated as,
\begin{equation}
auc_c \, = \, (auc_o -  a_{\hat{\pi}^\ast}) \, \frac{n_n + n_p + 1}{12 \, n_n \, n_p \, s_{\hat{\pi}^\ast}} + 0.5~.
\end{equation}

Furthermore, under the null hypothesis that the classifier has not learned the confounding signal, it follows that the confounding null distribution for the test statistic $\bar{AUC}^\ast = b^{-1} \sum_{i=1}^{b} AUC_i^\ast$ is given by,
\begin{equation}
N\left(\frac{1}{2} \, , \, \frac{n_n + n_p + 1}{12 \, n_n \, n_p \, b}\right)~.
\label{eq:meanconf}
\end{equation}

\section{Simulation experiments}

Here, we investigate the statistical properties of the response learning\footnote{Note that while Rao et al (2017) employed this test before, the authors did not evaluate their empirical performance.} ($H_0^\ast$ vs $H_1^\ast$) and confounding learning ($H_0^c$ vs $H_1^c$) statistical tests. The goal is to evaluate whether these are reasonably powered to detect response and confounding signals when those are present, and to evaluate the type I error rates achieved by the tests when data is simulated under the respective null hypothesis. But, before we describe in detail the simulation settings adopted in our experiments, we first explain how we simulated data from a binary classification task influenced by confounders.

\subsection{Synthetic data generation for binary classification}

We simulate data according to the model in Figure \ref{fig:strataucillustrations}a, where $C$ represents a binary confounder, $Y$ represents the disease status, and $X_1$, \ldots, $X_{10}$ represent the features.

This model is motivated by the following example involving a Parkinson's disease (PD) classification problem. Suppose that the data is unbalanced with respect to the gender ($C$) of the participants, with most of the male participants having PD and most of the female participants being controls (so that gender is strongly associated with the disease status). Because males on average are taller than females, it is reasonable to expect that many of the features extracted from the raw accelerometer data will likely be influenced by gender since taller participants tend to have larger step sizes, and different heel strikes and acceleration patterns than shorter participants (hence, we have $C \rightarrow X_j$). Furthermore, because PD patients experience difficulty to walk, their acceleration patterns tend to be distinct from control participants (hence, we have $Y \rightarrow X_j$). Therefore, it is reasonable to expect that the features will be influenced by disease status as well.

In order to generate an association between $C$ and $Y$ (i.e., $C \leftrightarrow Y$) we jointly sample these binary variables from a bivariate Bernoulli distribution (Dai, Ding, and Wahba 2013) \cite{dai2013}, with probability density function given by,
\begin{equation}
p(Y, C) \, = \, p_{11}^{y \, c} \, p_{10}^{y (1 - c)} \,  p_{01}^{(1 - y) c} \, p_{00}^{(1 - y) (1 - c)}~,
\end{equation}
where $p_{ij} = P(Y = i, C = j)$, and $p_{11} + p_{10} + p_{01} + p_{00} = 1$. Note that the covariance between $Y$ and $C$ is given by,
\begin{equation}
Cov(Y, C) \, = \, p_{11} \, p_{00} \, - \, p_{01} \, p_{10}~,
\end{equation}
and we can tune the strength of the association between $Y$ and $C$ by changing these parameters. Once, we have sampled a $\{y, c\}$ pair from this distribution, we sample the features from a multivariate normal distribution,
\begin{equation}
N_{10}\big( (y \, \beta + c \, \theta) \bf{1} \, , \, \bf{\Sigma} \big)~,
\end{equation}
where $\bf{1}$ represents the vector of ones, $\beta$ and $\theta$ are the regression coefficients, and $\bf{\Sigma}$ represents a correlation matrix with $ij$th element given by $\rho^{|i - j|}$.

\subsection{Parameter settings employed in the simulation experiments}

We performed four simulation experiments based on data generated with: disease and confounding signal ($H_1^\ast$ and $H_1^c$); disease but no confounding signal ($H_1^\ast$ and $H_0^c$); no disease or confounding signal ($H_0^\ast$ and $H_0^c$); and confounding but no disease signal ($H_0^\ast$ and $H_1^c$). In each experiment we generated 1,000 data sets. Each data set was generated (according to the model described above) using a unique combination of simulation parameter values. Table 1 presents the ranges of the simulation parameter values employed in each experiment.

\vspace{0.1cm}

\begin{table}[!h]
\begin{tabular}{lccccc}
\hline
& experiment 1 & experiment 2 & experiment 3 & experiment 4 \\
parameter & $H_1^\ast$, $H_1^c$ & $H_1^\ast$, $H_0^c$ & $H_0^\ast$, $H_1^c$ & $H_0^\ast$, $H_0^c$ \\ \hline
$n$ & \{200, \ldots, 600\} & \{200, \ldots, 600\} & \{200, \ldots, 600\} & \{200, \ldots, 600\} \\
$p_{11}$ & $[0.40 \, , \, 0.45]$ & $[0.40 \, , \, 0.45]$ & $[0.40 \, , \, 0.45]$ & $[0.40 \, , \, 0.45]$ \\
$p_{00}$ & $[0.40 \, , \, 0.45]$ & $[0.40 \, , \, 0.45]$ & $[0.40 \, , \, 0.45]$ & $[0.40 \, , \, 0.45]$ \\
$p_{10}$ & $[0.050 \, , \, 0.075]$ & $p_{11}$ & $[0.050 \, , \, 0.075]$ & $p_{11}$ \\
$p_{01}$ & $1-p_{11}-p_{00}-p_{10}$ & $p_{00}$ & $1-p_{11}-p_{00}-p_{10}$ & $p_{00}$ \\
$\beta$ & $[0.1 \, , \, 1.0]$ & $[0.1 \, , \, 1.0]$ & 0 & 0 \\
$\theta$ & $[0.1 \, , \, 1.0]$ & 0 & $[0.1 \, , \, 1.0]$ & 0 \\
$\rho$ & $[0.2 \, , \, 0.8]$ & $[0.2 \, , \, 0.8]$ & $[0.2 \, , \, 0.8]$ & $[0.2 \, , \, 0.8]$ \\
\hline
\end{tabular}
\caption{Ranges of the simulation parameter values employed in each experiment. Note that in experiments 2 and 4, the probs $p_{11}$, $p_{10}$, $p_{01}$, $p_{00}$ are re-normalized to sum 1.}
\end{table}

In order to select parameter values spread as uniformly as possible over the entire parameter range we employed a Latin hypercube design (Santner, Williams and Notz 2003), optimized according to the maximin distance criterion (Johnson, Moore and Ylvisaker 1990), in the determination of the parameter values used in the generation of each synthetic data set.

\subsection{Simulation results}

\begin{figure}[h]
  \centering
  \centerline{\includegraphics[width=\linewidth, bb = 0 0 790 180]{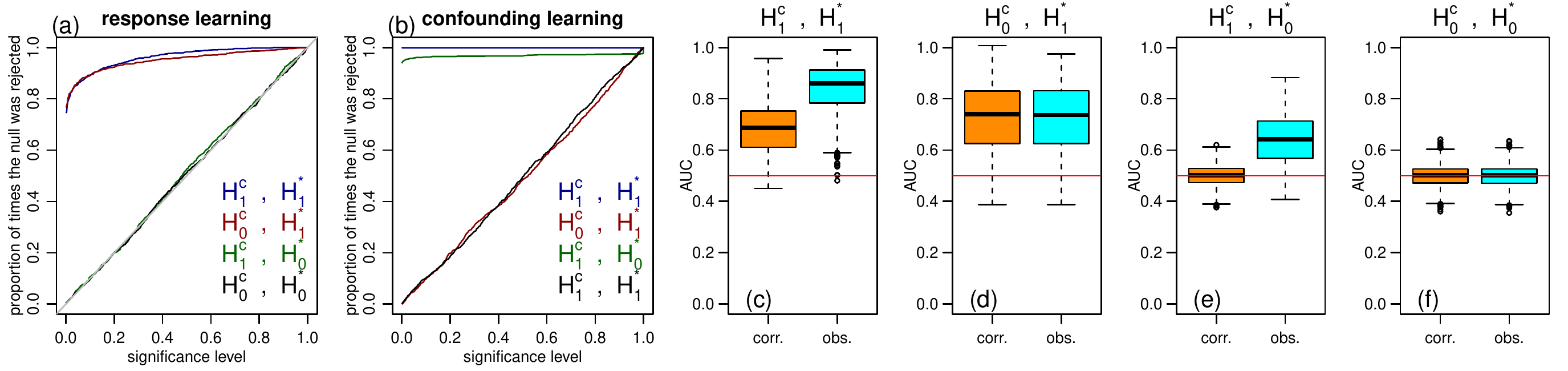}}
  \caption{Simulation study. The corrected AUCs and confounding tests were based on the analytical results described in Section 2.4. In each simulation, the number of permutations was set to equal the test set size.}
  \label{fig:simstudy}
\end{figure}

Figure \ref{fig:simstudy} reports the results from the four simulation experiments/settings described above (namely, $H_1^\ast$ and $H_1^c$;  $H_1^\ast$ and $H_0^c$; $H_0^\ast$ and $H_1^c$; and $H_0^\ast$ and $H_0^c$). In each experiment we generated 1,000 data sets and recorded the proportion of times that we rejected the null hypothesis across a grid of nominal significance levels varying from 0 to 1. Note that this proportion represents the empirical type I error rate when the data is simulated under the null, but the empirical power when the data is simulated under the alternative.

Panel a reports the power/error rate curves for the $H_0^\ast$ vs $H_1^\ast$ test. The blue and red curves represent the experiments with data simulated under $H_1^\ast$, and show that the restricted permutation test is well powered to detect disease learning in the presence (blue) or absence (red) of confounding. The green and black curves represent the simulation experiments under $H_0^\ast$ and show that the error rates are well controlled at the nominal significance levels in the presence (green) or absence (black) of confounding. (Note that a type I error rate curve close to the diagonal line implies that the distribution of the p-values is close to a uniform distribution in the 0 to 1 interval.)

Panel b reports the same results for the $H_0^c$ vs $H_1^c$ test. Again we observe well controlled errors under $H_0^c$ in the presence (red) and absence (black) of disease signal, and good power to detect confounding under $H_1^c$.

Finally, we also evaluated if our correction method was working as expected. Panels c-f show the distributions of the observed (cyan) and corrected (orange) AUCs for each of the 4 simulation experiments. In the presence of confounding (panels c and e), the corrected AUCs were lower than the observed AUCs, while in the absence of confounding (panels d and f), the corrected AUCs closely matched the observed values. Note, as well, that under $H_0^\ast$ the corrected AUCs were distributed around 0.5 (panels e and f), while the observed AUCs were still above of 0.5 in panel e due to confounding.

\section{Additional remarks: accounting for the confounder/response association structure in the population of interest}

In the main text, we define confounder as any variable causing spurious associations between the features and the response variable. We point out, however, that alternative definitions have also been used. For instance, Rao \textit{et al.} (2017) adopted a working definition where a covariate is considered to be a confounder if it is associated with the features and response and, additionally, if the joint distribution of the covariate and the response is shifted in the population-of-interest relative to the population used to develop the learner. Here, we describe how our tools can still be used if one is willing to adopt this definition.

In order to account for the confounder/response association structure in the population of interest we need to derive a baseline null distribution that preserves the structure observed in the population of interest, and use this distribution in place of the standard permutation null in our tools. For concreteness, we present next a detailed synthetic data example describing the approach.

\begin{figure}[!h]
  \centering
  \centerline{\includegraphics[width=4in, bb = 0 40 650 320]{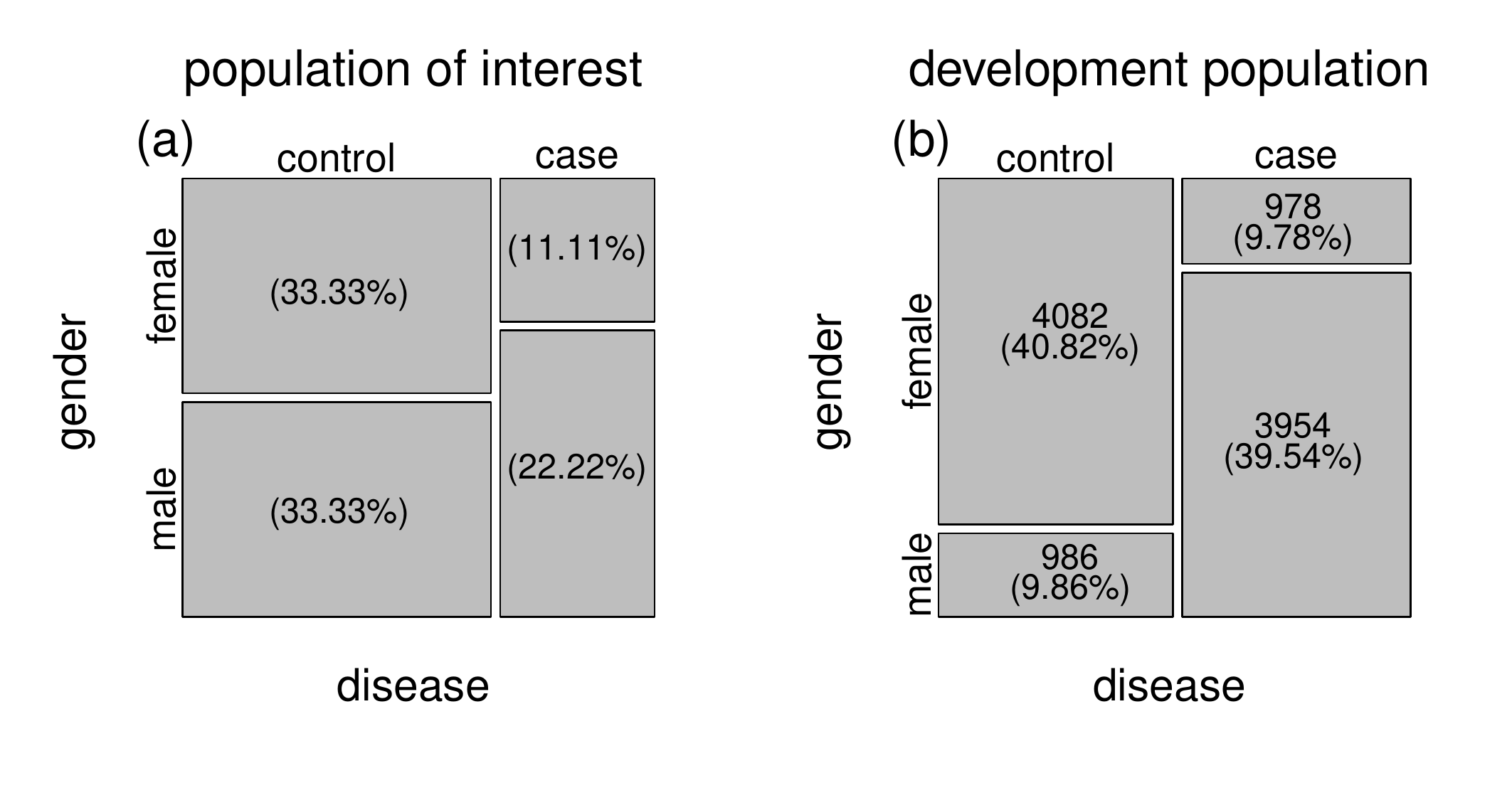}}
  \caption{Panel a describes the joint distribution between gender and disease status in the population of interest. This distribution was inferred from prior knowledge that the disease affects one third of the population and is two times more common in men than in women. Panel b shows the gender/disease status joint distribution actually observed in the synthetic data set (emulating a biased mobile health study) that will be used to develop a classifier for the disease. The numbers represent the observed absolute frequency in each gender/disease status class, while the porcentage of participants in each class are shown between parenthesis.}
  \label{sfig:mosaic1}
\end{figure}

Suppose we know that in the population of interest, a given disease affects one third of the population, and that the disease is two times more common in males than in females. The mosaic plot in Figure \ref{sfig:mosaic1}a describes the joint distribution of gender and disease status in the population of interest. Now, suppose that a mobile health study enrolled 10,000 participants, and we are interested in building a classifier of disease status. Suppose, as well, that due to self-selection mechanisms it turns out that gender and disease status are more strongly associated in the mobile health study data (i.e., the development population) than in the population of interest, as shown by the mosaic plot in Figure \ref{sfig:mosaic1}b (generated from synthetic data simulated with strong association between gender and disease status, as described in Section 3.1). Clearly, the development population is biased with respect to the population of interest in this example.

\begin{figure}[!h]
  \centering
  \includegraphics[width=2.7in]{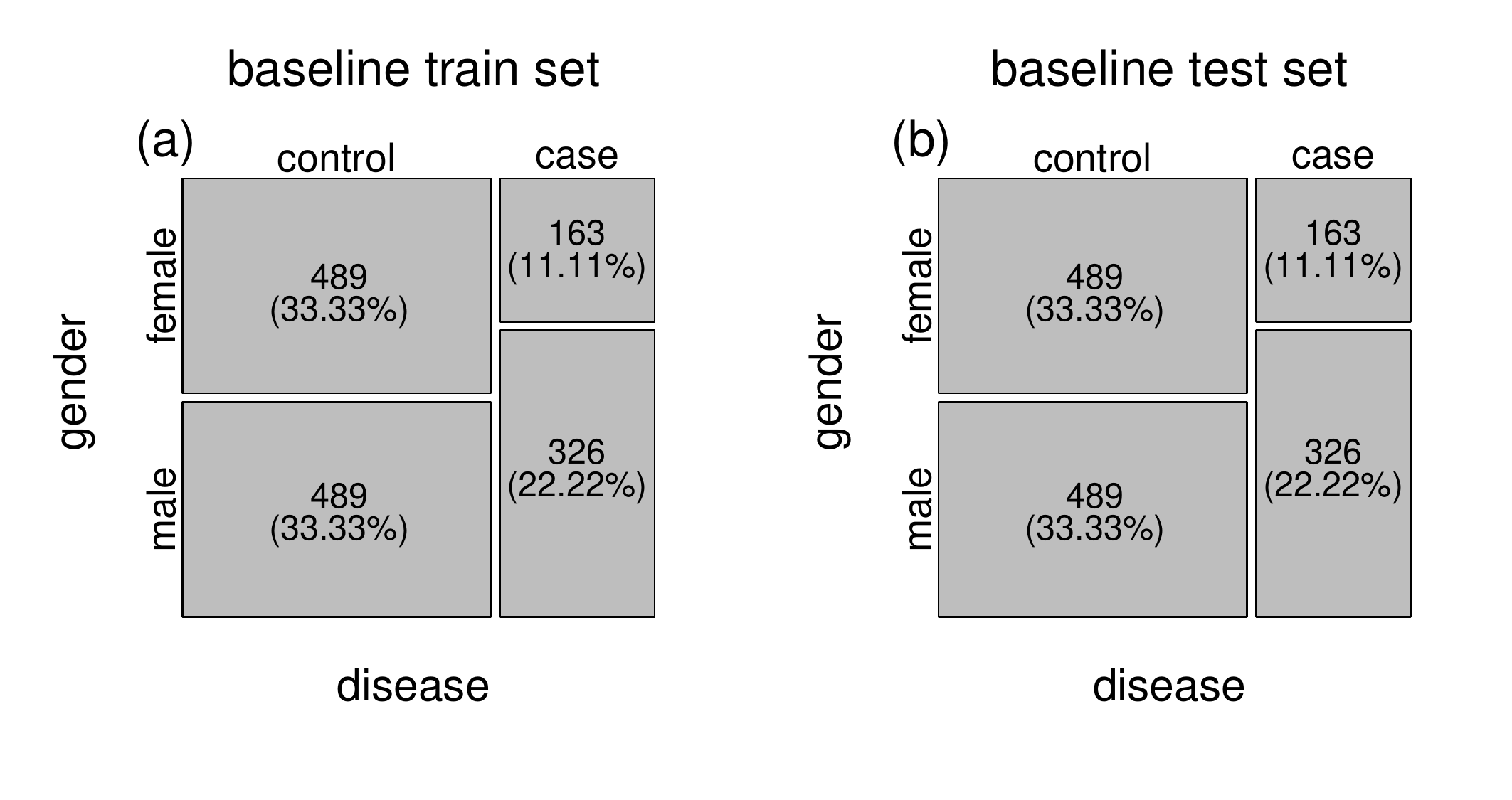}
  \includegraphics[width=2.7in]{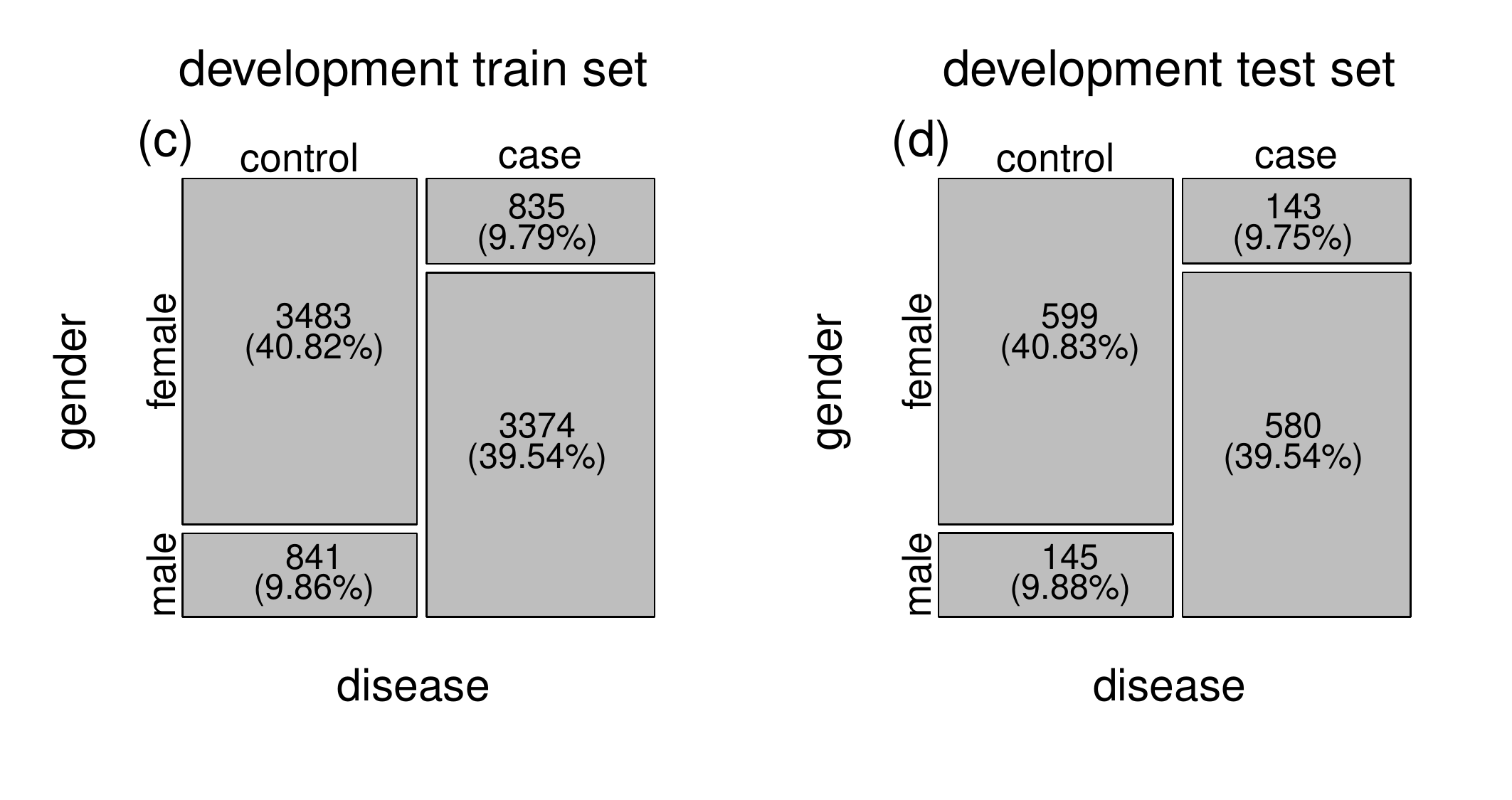}
  \vspace{-0.5cm}
  \caption{Panels a and b show the gender/disease status joint distribution for the baseline training and test sets. Note that these sets correspond to carefully selected sub-samples of the biased development data, that share the same gender/disease status joint distribution with the population of interest (Figure \ref{sfig:mosaic1}a). Panels c and d show the joint distribution for the training and test sets generated by splitting the biased development data. Note that these samples share the same gender/disease status joint distribution with the development data (Figure \ref{sfig:mosaic1}b). Observe, as well, that the development test set (panel d) has the same number of samples (namely, 1467) as the baseline test set (panel b).}
  \label{sfig:mosaic2}
\end{figure}

\begin{figure}[!h]
  \centering
  \centerline{\includegraphics[width=\linewidth]{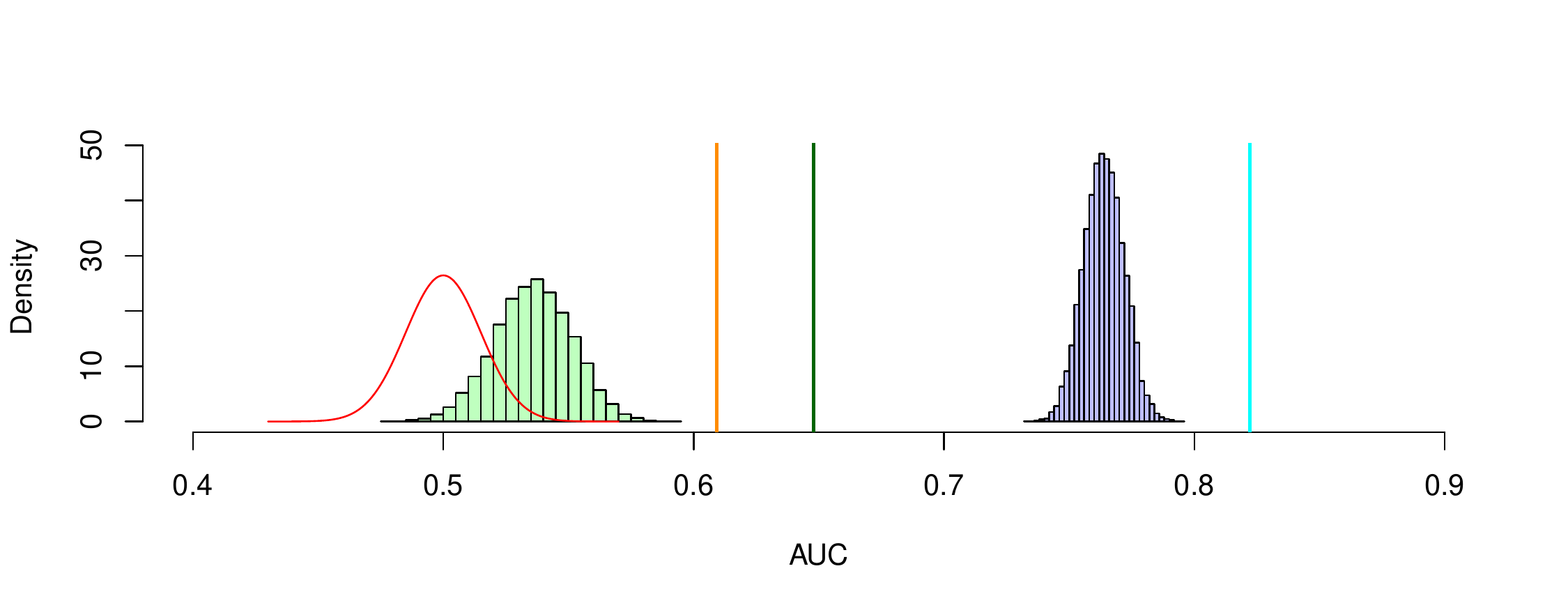}}
  \vspace{-0.5cm}
  \caption{The green histogram represents the baseline null distribution that captures the gender/response association structure in the population of interest. This baseline null was generated using restricted permutations applied to the baseline training and test sets described in Figure \ref{sfig:mosaic2}a and b. The blue histogram represents the restricted permutation null distribution generated from the development training and test sets described in Figures \ref{sfig:mosaic2}c and d. The red curve represents the density of the standard permutation null distribution. As before, the cyan line represents the uncorrected AUC value, while the orange line shows the corrected AUC (relative to the standard permutation null). The green line, on the other hand, represents the corrected AUC value relative to the population of interest.}
  \label{sfig:permdistrs}
\end{figure}

In order to account for the confounder/response association structure in the population of interest, we first need to generate a baseline null distribution that preserves this structure. To this end, we first sub-sample (from the development population) a training and test set showing the same joint distribution of gender and disease status as the population of interest. Figures \ref{sfig:mosaic2}a and b show the mosaic plots for these baseline training and test sets. Next, we apply restricted permutations to these subsets in order to generate the baseline null distribution (green histogram in Figure \ref{sfig:permdistrs}), which captures the gender/response association structure in the population of interest. (Note how this null distribution is shifted away from 0.5, due to the association between gender and disease status.)

In order to quantify the amount of confounding observed in the biased development data (relative to the population of interest) we first need to generate the restricted permutation null distribution for the biased data. To this end, we split the development data into training and test sets that preserve the joint distribution of the gender/disease label observed Figure \ref{sfig:mosaic1}b. (The test set, however, should have the same size as the baseline test set used to generate the baseline null (green histogram), in order to make these null distributions comparable.) Figures \ref{sfig:mosaic2}c and d show the mosaic plots for the training and test sets, while the blue histogram in Figure \ref{sfig:permdistrs} shows the restricted permutation null derived from the biased development data.

Now, in order to compute the corrected predictive performance of the classifier (relative to the population of interest) we only need to use the baseline null distribution in place of the standard permutation null. For instance, setting $a$ and $s$ to represent the mean and standard deviation of the baseline null (green histogram in Figure \ref{sfig:permdistrs}), and letting $a_{\hat{\pi}^{\ast}}$ and $s_{\hat{\pi}^{\ast}}$ represent, as before, the mean and standard deviation of the restricted permutation null (blue histogram in Figure \ref{sfig:permdistrs}), we can calculate the corrected AUC as,
\begin{equation}
(AUC_o - a_{\hat{\pi}^{\ast}}) \frac{s}{s_{\hat{\pi}^{\ast}}} + a~.
\end{equation}
The green line in Figure \ref{sfig:permdistrs} illustrates the above correction (while, for the sake of comparison, the orange one shows the correction with respect to the standard null distribution). (Note that the above correction formula corresponds to formula (\ref{eq:correctedmetric}), with $a$ and $s$ in place of $a_{\hat{\pi}^{\ast\ast}}$ and $s_{\hat{\pi}^{\ast\ast}}$.)

Similarly, we can still test for the presence of confounding (which in this example is measured by the amount to association between the features and response that goes beyond the association present in the population of interest). To this end, we can use the $N(a , s^2/b)$ distribution as an approximate null and compute the p-value for the confounding test as,
\begin{equation}
1 - \Phi\left( \frac{a_{\hat{\pi}^{\ast}} - a}{s/\sqrt{b}} \right)~.
\end{equation}

\section{Discussion}

In this paper we define confounder as any variable causing spurious associations between the features and the response variable. Our definition matches the one employed in causal analysis, where the goal is to estimate the causal effect of a variable on another. In the context of predictive modeling, a causality based definition has been adopted before by Linn et al. (2016). We point out, however, that alternative definitions have also been used. For instance, Rao et al. (2017) adopted a working definition where a covariate is considered to be a confounder if it is associated with the features and response and, additionally, if its joint distribution with the response is shifted in the population-of-interest relative to the population used to develop the model. In Section 4 above, we describe how our tools can still be used if one is willing to adopt this definition.

Our confounding tests provide a principled and direct approach to check if potential confounders are truly influencing the predictive performance of a learner. Recall that demographic variables associated with the response, but not with the features, will not confound the prediction. Hence, the proposed tests can be used to assess which potential confounders are really problematic. If an important confounder is detected, then the novel correction approach propose in this paper can be used to estimate how much the algorithm has actually learned about the response. Observe, however, that this correction is done ``after the fact", when the algorithm has already had a chance to learn both response and confounding signals. (Hence, it does not correspond to an usual confounding adjustment method, in the sense that it does not prevent an algorithm from learning the confounding signal in the first place. It simply quantifies the actual amount of response signal learned by the algorithm.) Ideally, one should try to prevent an algorithm from learning the confounding signal in the first place. Most importantly, once our tests show that a potentially problematic confounder is indeed confounding the response/feature association, and we attempt to adjust for the confounder using any confounding adjustment method (where confounding adjustment should be understood as avoiding that the algorithm learns the confounder signal to begin with), then our tests can be used again to evaluate if the confounding adjustment method has worked or not.

In a previous contribution (Chaibub Neto \textit{et al.} 2017), we proposed permutation tests to detect disease signal in the presence of ``identity confounding", as well as, permutation tests to detect the ``identity confounding" itself. The ``identity confounding" issue arises when the relationship between the features and the disease labels learned by a classifier is confounded by the identity of the subjects. This problem can arise when longitudinal records provided by each participant are split into the training and test sets (record-wise data split), rather than being assigned to either the training or the test sets (subject-wise data split). When data from each participant is present in both the training and test sets, the classifier might be mostly learning about the subject's individual characteristics rather than the disease signature. Detecting identity confounding, however, requires a different permutation strategy (i.e., ``subject-wise" label shufflings), rather than the restricted permutations adopted in this paper.

In Section 3 we report a simulation study evaluating the statistical properties of the $H_0^\ast$ vs $H_1^\ast$ and $H_0^c$ vs $H_1^c$ tests. For both tests, we observed reasonable statistical power for the range of parameters investigated in our experiments. We also observed that the type I error rates were well controlled and close to the nominal significance levels, suggesting that the permutation tests were close to exact. (Note that the only assumption required to guarantee the exactness of the restricted permutation test is that, under $H_0^\ast$, the response data is exchangeable within each level of the confounder variable. The test does not require any distributional assumptions and is applicable to any performance metric in both regression and classification problems. The confounding test, on the other hand, makes the additional assumption that the test set is large.)

To the best of our knowledge, the use of restricted permutations for testing response learning in the presence of confounding has only been leveraged by Rao et al. (2017). These authors, however, did not take the next logical steps of developing statistical tests to detect confounding per se, or correction formulas to estimate the amount of response signal learned by the algorithm that might also have learned confounding signal.

Finally, we point out that while this paper has focused in mobile health applications, the proposed tools can be more generally applied to any other areas plagued by confounders.

\section{Appendix}

\subsection{Proof of Theorem 1}

Let's start with the covariance operator. By definition,
\begin{equation}
pCov(X, Y \mid C) \, = \, Cov(X, Y) \, - \, Cov(X, C) \, Var(C)^{-1} \, Cov(Y, C)~.
\label{eq:pcovdef}
\end{equation}

Hence,
\begin{equation}
pCov(X, Y^\ast \mid C) \, = \, Cov(X, Y^\ast) \, - \, Cov(X, C) \, Var(C)^{-1} \, Cov(Y^\ast, C)~,
\end{equation}
and
\begin{equation}
E_{\pi^\ast}\left[pCov(X, Y^\ast \mid C)\right] \, = \, E_{\pi^\ast}\left[Cov(X, Y^\ast)\right] \, - \, E_{\pi^\ast}\left[Cov(X, C) \, Var(C)^{-1} \, Cov(Y^\ast, C)\right]~,
\end{equation}
where the expectation is taken with respect to the restricted permutation null distribution, $F_{\pi^\ast}$.

Now, recall that $Cov(X, Y)$ measures the total amount of association between $X$ and $Y$, which can be partitioned into a the component due to the direct association between $X$ and $Y$ (due to a potential causal relation between these two variables, or to the presence of additional unmeasured confounders associated with $X$ and $Y$) and into an indirect component where part of the association between $X$ and $Y$ is explained by the association between $X$ and $C$, and between $C$ and $Y$. Because the restricted shuffling of $Y$ with the respect to the levels of $C$ preserves the associations between $Y$ and $C$ and between $X$ and $C$, we have that the restricted shuffling removes the direct association between $X$ and $Y$, leaving intact the indirect association component mediated by $C$. In other words, the average association between $X$ and $Y^\ast$ (across all possible restricted permutations of $Y$) is completely explained by $C$. Hence, it follows that conditional on $C$, this average association vanishes, that is,
\begin{equation}
E_{\pi^\ast}\left[pCov(X, Y^\ast \mid C)\right] \, = \, 0~,
\end{equation}
so that,
\begin{align}
E_{\pi^\ast}\left[Cov(X, Y^\ast)\right] \, &= \, E_{\pi^\ast}\left[Cov(X, C) \, Var(C)^{-1} \, Cov(Y^\ast, C)\right] \nonumber \\
&= \, E_{\pi^\ast}\left[Cov(X, C) \, Var(C)^{-1} \, Cov(Y, C)\right] \nonumber \\
&= \, Cov(X, C) \, Var(C)^{-1} \, Cov(Y, C)~, \label{eq:equality}
\end{align}
where the second equality follows from the fact that $Cov(Y^\ast, C) = Cov(Y, C)$. Therefore, it follows from (\ref{eq:pcovdef}) and (\ref{eq:equality}) that,
\begin{equation}
pCov(X, Y \mid C) \, = \, Cov(X, Y) \, - \, E_{\pi^\ast}\left[Cov(X, Y^\ast)\right]~.
\end{equation}

To prove the result for the partial correlation, observe that by dividing the above equation by $\sqrt{Var(X) Var(Y)}$, and recalling that by definition $Cor(X, Y) = Cov(X, Y)/\sqrt{Var(X) Var(Y)}$ we have that,
\begin{align}
\frac{pCov(X, Y \mid C)}{\sqrt{Var(X) Var(Y)}} \, &= \, Cor(X, Y) \, - \, E_{\pi^\ast}\left[\frac{Cov(X, Y^\ast)}{\sqrt{Var(X) Var(Y)}}\right]~, \\ \nonumber
&= \, Cor(X, Y) \, - \, E_{\pi^\ast}\left[Cor(X, Y^\ast)\right]~,
\end{align}
where the second equality follows from the fact that $Var(Y) = Var(Y^\ast)$. Now, recalling that by definition $pCor(X, Y \mid C) = Cov(X, Y \mid C)/\sqrt{Var(X \mid C) Var(Y \mid C)}$, the result follows by dividing and multiplying the left hand side of the above equation by $\sqrt{Var(X \mid C) Var(Y \mid C)}$ and rearranging terms.

\subsection{Theorem 2}

\setcounter{theorem}{1}
\begin{theorem}
Let $C$ represent a categorical variable, $X$ and $Y$ represent arbitrary random variables, and $Y^\ast$ represent a restricted permutation of $Y$ relative to the levels of $C$. Let $dCov(X, Y)$ represent the distance covariance between $X$ and $Y$, and $pdCov(X, Y \mid C)$ represent the partial distance covariance between $X$ and $Y$ given $C$. An alternative formula for the computation of $pdCov(X, Y \mid C)$, based on the restricted permutations of $Y$ with respect to the levels of $C$, is given by,
\begin{equation}
pdCov(X, Y \mid C) \, = \, dCov(X, Y)^2  \, - \, E_{\pi^\ast}\left[ dCov(X, Y^\ast)^2 \right]~.
\label{eq:pdcovperm}
\end{equation}
Similarly, the partial distance correlation $pdCor(X, Y \mid C)$ can be expressed as,
\begin{equation}
pdCor(X, Y \mid C) \, = \, \frac{dCor(X, Y)^2 - E_{\pi^\ast}\left[ dCor(X, Y^\ast)^2 \right]}{\sqrt{(1 - dCor(X, C)^4) \, (1 - dCor(Y, C)^4)}}~.
\label{eq:pdcorperm}
\end{equation}
\end{theorem}

\subsubsection{Proof of Theorem 2}

Consider first the partial distance covariance operator. By definition (see equation A.1 in Appendix A.5 of reference\cite{pdcor2014}),
\begin{equation}
pdCov(X, Y \mid C) \, = \, dCov(X, Y)^2 \, - \, dCov(X, C)^2 \, dVar(C)^{-2} \, dCov(Y, C)^2~.
\label{eq:pdcovdef}
\end{equation}
Following the same reasoning used in the proof of Theorem 1 we obtain equation (\ref{eq:pdcovperm}).

Now, let's consider the partial distance correlation. By definition the distance correlation is given by\cite{dcor2007},
\begin{equation}
dCor(X, Y) \, = \, \frac{dCov(X, Y)}{\sqrt{dVar(X) \, dVar(Y)}}~.
\end{equation}

Dividing equation (\ref{eq:pdcovperm}) by $dVar(X) \, dVar(Y)$ we have that
\begin{align}
\frac{pdCov(X, Y \mid C)}{dVar(X) \, dVar(Y)} \, &= \, dCor(X, Y)^2 \, - \, E_{\pi^\ast}\left[\frac{dCov(X, Y^\ast)^2}{dVar(X) \, dVar(Y)}\right]~, \\
&= \, dCor(X, Y)^2 \, - \, E_{\pi^\ast}\left[dCor(X, Y^\ast)^2\right]~,
\label{eq:pdcorderivation}
\end{align}
where the second equality follows from the fact that $dVar(Y) = dVar(Y^\ast)$.

By definition the partial distance correlation (see Appendix A.5 of reference\cite{pdcor2014}) is given by,
\begin{equation}
pdCor(X, Y \mid C) \, = \, \frac{pdCov(X, Y \mid C)}{\sqrt{dVar(X)^2 (1 - dCor(X, C)^4)} \sqrt{dVar(Y)^2 (1 - dCor(Y, C)^4)}}~.
\end{equation}

Hence, we obtain the result in equation (\ref{eq:pdcorperm}) by dividing and multiplying the left hand side of equation (\ref{eq:pdcorderivation}) by $\sqrt{dVar(X)^2 \, (1 - dCor(X, C)^4) \, dVar(Y)^2 \, (1 - dCor(Y, C)^4)}$ and rearranging terms.

\subsection{Synthetic data generation for correlation metric illustrations}

For the correlation illustrations, presented on Figure 2 in the main text, we simulated data from the model,
$$
\xymatrix{
 & *+[F-:<10pt>]{C} \ar[dl]_{\beta_{xc}} \ar[dr]^{\beta_{yc}} &  \\
*+[F-:<10pt>]{X} && *+[F-:<10pt>]{Y} \ar[ll]^{\beta_{xy}} \\
}
$$
where $C$ is a binary confounder sampled from a Bernoulli distribution with probability of success $p$; $Y$ is sampled from a $N(\beta_{yc} \, c \, , \, 1)$; and $X$ is sampled from a $N(\beta_{xc} \, c + \beta_{xy} \, y \, , \, 1)$. For each of the 1,000 simulations we used a distinct set of simulation parameters ($p$, $\beta_{xc}$, $\beta_{yc}$, and $\beta_{xy}$), randomly draw from uniform distributions with ranges,
$$
p \, \sim \, U(0.3 \, , \, 0.7)~, \hspace{0.3cm}
\beta_{xc} \, \sim \, U(-3 \, , \, 3)~, \hspace{0.3cm}
\beta_{yc} \, \sim \, U(-3 \, , \, 3)~, \hspace{0.3cm}
\beta_{yx} \, \sim \, U(-3 \, , \, 3)~,
$$
and sample size 1,000.

\subsection{Algorithm 2}

In order to test whether an algorithm is learning the confounding signal, we need to generate a permutation null distribution (for the $\bar{m}^\ast$ statistic) where the indirect association mediated by the confounder is destroyed. To this end, we shuffle the confounder vector in a standard fashion, before computing the $\bar{m}^\ast$ statistic, as described in Algorithm \ref{alg:doubleShuffling} below.
\setcounter{algorithm}{1}
\begin{algorithm}[H]
\caption{Monte Carlo permutation null distribution to detect confounding}\label{alg:doubleShuffling}
\begin{algorithmic}[1]
\State \textbf{Input}: Number of standard permutations, $b_s$; $\bfX$; $\bfy$; $\bfc$; training and test set indexes, $i_{train}$, $i_{test}$
\State Split $\bfX$, $\bfy$ and $\bfc$ into training and test sets
\State Set the number of restricted permutations to the test set size, $b_r \leftarrow \mbox{Length}(i_{test})$
\For{$i = 1, 2, \ldots, b_s$}
  \State $\bfc^{\ast\ast}_{train} \leftarrow \mbox{StandardShuffle}(\bfc_{train})$, $\bfc^{\ast\ast}_{test} \leftarrow \mbox{StandardShuffle}(\bfc_{test})$
  \For{$j = 1, 2, \ldots, b_r$}
    \State $\bfy^{\ast}_{train} \leftarrow \mbox{RestrictedShuffle}(\bfy_{train}, \bfc^{\ast\ast}_{train})$, $\bfy^{\ast}_{test} \leftarrow \mbox{RestrictedShuffle}(\bfy_{test}, \bfc^{\ast\ast}_{test})$
    \State Train a machine learning algorithm on the $\bfX_{train}$ and $\bfy_{train}^{\ast}$ data
    \State Evaluate the algorithm on the $\bfX_{test}$ and $\bfy_{test}^{\ast}$ data
    \State Record the value of the performance metric, $m^\ast_j$, on the shuffled data
  \EndFor
  \State Compute and store $\bar{m}^\ast_i = b_r^{-1} \sum_{j = 1}^{b_r} m^\ast_j$
\EndFor
\State \textbf{Output}: $\bar{m}^\ast_1$, $\bar{m}^\ast_2$, \ldots, $\bar{m}^\ast_{b_s}$
\end{algorithmic}
\end{algorithm}

\end{document}